\documentclass[aps,prd,twocolumn,superscriptaddress,nofootinbib,a4paper]{revtex4-2}

\usepackage{fontspec}
\usepackage[english]{babel}
\usepackage{amsmath,amssymb,mathtools,amsfonts}
\usepackage{physics,slashed}
\usepackage[compat=1.0.0]{tikz-feynman}
\usepackage{xcolor}
\usepackage{hyperref}
\hypersetup{hypertexnames=false}

\newcommand{\sbk}[1]{\left[{#1}\right]}
\newcommand{\abk}[1]{\left\langle{#1}\right\rangle}
\newcommand{\asbk}[1]{\left\langle{#1}\right]}

\begin{document}

\title{An Efficient On-shell Framework for EFT Matching}

\author{Ziyu Dong}
\email{zdong@ifae.es}
\affiliation{IFAE and BIST, Universitat Aut\`onoma de Barcelona, 08193 Bellaterra, Barcelona}

\author{Cihang Li}
\email{lch26@stu.pku.edu.cn}
\affiliation{School of Physics, Peking University, Beijing 100871, China}

\author{Teng Ma}
\email{mateng@ucas.ac.cn}
\affiliation{International Centre for Theoretical Physics Asia-Pacific (ICTP-AP), University of Chinese Academy of Sciences (UCAS), 100190 Beijing, China}

\author{Jing Shu}
\email{jshu@pku.edu.cn}
\affiliation{School of Physics and State Key Laboratory of Nuclear Physics and Technology, Peking University, Beijing 100871, China}
\affiliation{Center for High Energy Physics, Peking University, Beijing 100871, China}
\affiliation{Beijing Laser Acceleration Innovation Center, Huairou, Beijing, 101400, China}

\author{Zizheng Zhou}
\email{zhouzizheng@itp.ac.cn}
\affiliation{CAS Key Laboratory of Theoretical Physics, Institute of Theoretical Physics, Chinese Academy of Sciences, Beijing 100190, China}
\affiliation{School of Physical Sciences, University of Chinese Academy of Sciences, Beijing 100049, China}

\begin{abstract}
Standard techniques for one-loop EFT matching often have gauge and basis redundancies, while strictly 4-dimensional on-shell methods fail to capture rational terms. To resolve this, we develop an efficient, channel-based on-shell framework for one-loop matching. Our method reconstructs the local hard-region amplitude by sewing tree amplitudes across double cuts, employing a mass-shift prescription to recover d-dimensional internal states. This d-dimensional sewing retains rational terms and integrates them seamlessly with ordinary cut-constructible contributions. The local amplitude is expanded in the hard region and directly projected onto non-redundant, on-shell EFT amplitude bases. With tadpoles and kinematically independent bubbles systematically fixed by an explicit subtraction convention, our framework successfully merges the rigorous extraction of rational Wilson coefficients with the gauge-invariant elegance of modern amplitude methods.
\end{abstract}

\maketitle

\section{Introduction}
\label{sec:intro}

Effective field theories (EFTs) describe low-energy observables after heavy degrees of freedom have been integrated out.
At a matching scale $M$, the Wilson coefficients are fixed by the short-distance part of UV amplitudes; subsequent renormalization group evolution resums logarithms between separated scales.
At one loop, matching coefficients are extracted by performing a local hard-region expansion of the UV amplitude in inverse powers of the heavy scale $M$.

Standard Feynman-diagram matching computes this local expansion directly, but gauge fixing, ghosts, and redundant operator bases can obscure the physical structures that survive in the EFT.
The Covariant Derivative Expansion (CDE)~\cite{henningHowUseStandard2016,henningOneloopMatchingRunning2018,cohenFunctionalPrescriptionEFT2021} keeps gauge covariance manifest, but its output still has to be reduced to a non-redundant EFT basis.
On-shell amplitude bases provide a complementary organization: local on-shell contact amplitudes are built from physical external states and can be chosen to remove equations-of-motion and integration-by-parts redundancies~\cite{shadmiEffectiveFieldTheory2019,Ma2019gtx, henningConstructingEffectiveField2019, durieuxConstructingMassiveOnshell2020, dongConstructingGenericEffective2022,liCompleteSetDimension82021,Liu:2023jbq}.
This makes the final projection onto gauge-invariant EFT structures direct once the local UV amplitude is known.

The remaining issue is how to obtain the local UV amplitude from on-shell data at one loop.
Generalized unitarity methods~\cite{Britto:2004nc,Giele:2008ve,Badger:2008cm,Ellis:2007br} reconstruct loop amplitudes from cuts and have been used in EFT matching~\cite{bernGravitationalEffectiveField2021, delleroseWilsonCoefficientsNatural2022}.
Four-dimensional cuts, however, do not by themselves determine rational terms generated by the extra-dimensional loop momentum, and tadpoles or kinematically independent bubbles have no ordinary kinematic discontinuity.
Dispersion-based approaches provide another on-shell route to matching~\cite{DeAngelis:2023bmd}, with a different treatment of analytic information.

This paper adapts standard $d$-dimensional unitarity sewing and mass-shift reconstruction to one-loop on-shell EFT matching.
The main contribution is a practical matching prescription that organizes these ingredients into a single workflow.
We sew double cuts using $d$-dimensional unitarity with mass-shifted internal kinematics.
After reducing the channel results to a common scalar-integral basis, we apply channel projection and overlap removal.
Rational terms are assigned the same channel support as their parent d-dimensional scalar coefficients before the $\epsilon \to 0$ limit.
To handle tadpoles and kinematically independent bubbles (KIBs), we use an on-shell-like subtraction scheme.
At the two-point level, the difference from a standard on-shell scheme is a finite field redefinition; in amplitudes with interactions this induces the corresponding finite redefinition of renormalized parameters before the Wilson coefficients are read off.

We work at one loop, with four-dimensional external states and $d$-dimensional internal loop reconstruction.
Once the one-loop amplitude is reconstructed and reduced to scalar masters, the Wilson coefficients are obtained by applying the hard-region expansion and projecting the resulting local amplitude onto the EFT basis.
The paper is organized as follows.
Sec.~\ref{sec:framework} defines the sewing, projection, rational-term, and subtraction prescriptions.
Sec.~\ref{sec:matching_procedure} explains how the reduced one-loop result is expanded in the hard region and projected onto on-shell EFT bases.
We then demonstrate this framework through explicit examples: Sec.~\ref{sec:sqed} uses the sQED four-photon amplitude as an amplitude-level rational reconstruction benchmark, and Sec.~\ref{sec:massive_vector} uses a massive-vector model to show how the rational term can drive the local loop contribution.
Appendix~\ref{app:dimensional-conventions} fixes the dimensional and spinor-helicity conventions and gives the practical $\tilde\mu$ reconstruction used for tree amplitudes entering the cuts.
Appendix~\ref{sec:appendix-matching-4Fermion} presents a mixed massive--massless four-fermion matching calculation, Appendix~\ref{sec:appendix-scalar-integral-convention} defines the scalar-integral normalization, Appendix~\ref{sec:appendix-hard-region} lists the hard-region expansions used to extract the Wilson coefficients, and Appendix~\ref{App:rational} explains how rational parts enter the EFT matching examples.

\section{The Framework of D-Dimensional Unitarity Sewing}
\label{sec:framework}

The purpose of the construction is to obtain the local hard-region part of a one-loop UV amplitude from on-shell data.
This requires a $d$-dimensional loop integrand before the $\epsilon\to0$ limit, because rational terms and finite local terms associated with tadpoles and KIBs can enter the Wilson coefficients.

\subsection{Cut kinematics and channel projection}
\label{sec:algorithm-assumptions}

The construction in this section is organized around the one-loop amplitude data that will later be expanded in the hard region and matched onto the EFT basis.
After tensor reduction, a one-loop amplitude can be written in the standard scalar-integral form~\cite{gerardthooftScalarOneLoop1979,chetyrkin1981integration,bernDimensionallyRegulatedOneloop1993,dennerReductionOneloopTensor2003}
\begin{equation} \label{eq:scalar-integrals-basis-d4}
\mathcal{A}_N^{(1)}=\sum_{a=1}^{4}\sum_i C_a^{\{i\}} I_a^{\{i\}}+R\,,
\end{equation}
where $a=4,3,2,1$ labels boxes, triangles, bubbles, and tadpoles, respectively.
The symbol $I_a^{\{i\}}$ denotes the scalar integral for the corresponding topology and kinematic routing, $C_a^{\{i\}}$ is its coefficient, and $R$ is the rational term.
This decomposition separates the reconstruction problem into three parts: branch-cut scalar-integral coefficients, the rational remainder, and local tadpole or KIB terms without physical branch cuts.

Four-dimensional unitarity cuts determine the coefficients of scalar integrals with physical branch cuts, such as boxes, triangles, and kinematically dependent bubbles.
The rational term $R$ can receive contributions from the $(-2\epsilon)$-dimensional components of loop momenta and internal-state algebra.
Tadpoles and kinematically independent bubbles (KIBs), which have no physical branch cut, also require a subtraction prescription.
Since each sector can feed the local hard-region amplitude, the cut calculation has to retain the dimensional loop information before the $\epsilon\to0$ limit.

This requirement fixes the division of kinematics used throughout the paper.
The final EFT projection is written with four-dimensional external helicity states, so all external particles are kept as four-dimensional on-shell states with physical polarizations and four-dimensional spinor-helicity variables.
The cut loop momenta and the algebra of internal states are treated in dimensional regularization.

In the BMHV convention summarized in Appendix~\ref{app:dimensional-conventions}, a loop momentum is decomposed as $\bar l=l+\vec{\tilde\mu}$, where $l$ is four-dimensional and $\vec{\tilde\mu}$ lies in the $(-2\epsilon)$-dimensional subspace.
We use the sign convention $\bar l^2=l^2-\tilde\mu^2$, with $l\cdot\vec{\tilde\mu}=0$ and $\tilde\mu^2\equiv-\vec{\tilde\mu}^{\,2}$ for the extra-dimensional component.
For an internal line of physical mass $m$, the $d$-dimensional cut condition $\bar l^2=m^2$ is represented in the four-dimensional spinor variables by
\begin{equation}
    l^2=m^2+\tilde\mu^2 \equiv m_*^2 .
\end{equation}
The shifted mass $m_*$ is therefore the four-dimensional representation of the internal $d$-dimensional on-shell condition.
It enters the on-shell conditions, spin and polarization sums, gamma traces, and contractions involving cut loop momenta.
External masses, Lagrangian parameters, Wilson coefficients, and couplings keep their physical definitions.

With these kinematic rules fixed, the one-loop amplitude is reconstructed channel by channel from double cuts.
For a given channel, two on-shell tree amplitudes are sewn over the internal states, and the cut conditions are then lifted back to Feynman propagators.
The resulting expression carries the branch cut of the selected channel, and multi-channel topologies can generate scalar integrals shared with other channels.
The sewn channel expressions are therefore reduced to a common scalar-integral basis before channel content is assigned.

The channel-selection prescription $\mathcal P_{s_k}$ keeps the scalar integrals whose discontinuity contains the channel $s_k$, together with the rational contributions associated with the same $d$-dimensional parent coefficients.
Overlap among multi-channel scalar integrals is then removed or weighted according to Eq.~(\ref{eq:merging_abstract}).
After the reduced UV amplitude has been reconstructed, the matching step keeps its hard-region local part.
The hard-region expansion, performed either at the integrand level or on the scalar master integrals after tensor reduction, expands around loop momentum of order the heavy scale, $\bar l\sim M\gg p_i,m_i$.
The soft-region contribution reproduces the EFT loop contribution in the common infrared prescription, while the hard-region polynomial is projected onto the on-shell EFT amplitude basis to read off Wilson coefficients.
Tadpoles and kinematically independent bubbles are fixed by the local subtraction convention of Sec.~\ref{sec:decoupling}; changing that convention amounts to a finite local scheme transformation.

The rest of the framework follows this organization.
We first sew double cuts to reconstruct the branch-cut scalar-integral data, then use the mass-shifted internal-state reconstruction to retain the rational contributions associated with the same $d$-dimensional parent terms.
The four-photon amplitude in massless sQED provides an amplitude-level benchmark for this rational reconstruction, and the massive-vector example shows a simple matching calculation in which the rational term gives the leading displayed local loop contribution.
The remaining KIB and tadpole terms are fixed by the subtraction prescription of Sec.~\ref{sec:decoupling}.

\subsection{One-loop amplitudes via unitarity sewing}
\label{sec:unitarity_cuts}

The unitarity of the S-matrix relates the discontinuity of a one-loop amplitude across a branch cut to the product of two on-shell tree-level amplitudes.
This property allows us to reconstruct the loop integrand involving $C_4$, $C_3$, and kinematically dependent $C_2$ terms without introducing unphysical degrees of freedom such as ghosts or gauge-fixing terms.

Full generalized unitarity could also be used to isolate the master-integral coefficients directly through a hierarchy of multiple cuts.
For the one-loop matching examples considered here, the double-cut channel organization is sufficient and economical: after tensor reduction to a common scalar-integral basis, the parent scalar integrals and their rational descendants can be assigned without introducing the heavier machinery of multiple-cut coefficient extraction.
The two organizations use the same underlying $d$-dimensional unitarity information; the choice made here is a channel organization tailored to the present matching calculations.
Specifically, in the four-dimensional cut notation used for the cut phase space, the discontinuity across a kinematic channel $s_\mathcal{I}$ is obtained by sewing two tree amplitudes $\mathcal{A}_L$ and $\mathcal{A}_R$ along the cut loop propagators:
\begin{equation} \label{eq:disc}
    \begin{aligned}
        &-i\left.\operatorname{Disc} \mathcal{A}^{(1)} \right|_{s_\mathcal{I}}
        \\&= \sum_{\sigma_1, \sigma_2}  \int \frac{\dd^4 l_1}{(2 \pi)^{2}}
         \times \delta^+\left(l_1^2-m_1^2\right) \delta^+\left(l_2^2-m_2^2\right)
        \\& \times \mathcal{A}^{(0)}_L\left(l_2^{\sigma_2}, \dots, l_1^{\sigma_1}\right)
        \times \mathcal{A}^{(0)}_R\left(-l_1^{-\sigma_1}, \dots, -l_2^{-\sigma_2}\right),
    \end{aligned}
\end{equation}

where $\delta^+(q^2-m^2)\equiv \theta(q^0)\delta(q^2-m^2)$ enforces the positive-energy on-shell condition, $s_\mathcal{I}=(l_1+l_2)^2$, and $l_{1,2}$ are four-dimensional representatives of the cut propagators.
Equation~(\ref{eq:disc}) should therefore be read as the channel-sewing formula before the extra-dimensional information is restored.
For the rational reconstruction used in this paper, the internal on-shell condition is the BMHV one, $\bar l_i^2=m_i^2$, represented by $l_i^2=m_i^2+\tilde\mu_i^2$ in four-dimensional spinor variables, and the closed internal-state algebra is evaluated in $d=4-2\epsilon$ dimensions as described in Sec.~\ref{sec:rational-recovering}.

To reconstruct the full quantum amplitude, the delta functions must be replaced by standard Feynman propagators.
However, a Feynman propagator contains a principal value part in addition to the on-shell condition, as given by $1/(x \pm i\epsilon) = \mathcal{P}(1/x) \mp i\pi\delta(x)$.
Therefore, applying the reverse replacement alone
\begin{equation}\label{eq:onshell-to-prop}
    2\pi\delta^+(l^2-m^2) \rightarrow \frac{i}{l^2-m^2+i\epsilon}
\end{equation}
produces redundant terms that possess branch cuts outside the target $s_\mathcal{I}$-channel.
These redundancies must be removed consistently during the amplitude assembly, as illustrated later in Eq.~(\ref{eq:merging_abstract}).

In a strictly four-dimensional cut, the numerator of the loop integrand is reconstructed by summing over the internal physical states.
We use four-dimensional spinor-helicity conventions following Ref.~\cite{ochirov2018massiveqcd}.
For a massless momentum, $p_{\alpha\dot\alpha}=|p\rangle_\alpha[p|_{\dot\alpha}$ and $\langle pq\rangle[qp]=2p\cdot q$.
For a massive four-dimensional momentum $p^2=m^2$, the spinors carry an $SU(2)$ little-group index $I=1,2$, raised and lowered by the antisymmetric tensor, and satisfy
\begin{equation}\label{eq:completeness}
    \begin{aligned}
        |p^I\rangle_\alpha[p_I|_{\dot\alpha}
        &=p_{\alpha\dot\alpha}\,,\\
        |p^I\rangle_\alpha\langle p_I|^\beta
        &=-m\delta_\alpha^{\ \beta}\,,\\
        |p^I]^{\dot\alpha}[p_I|_{\dot\beta}
        &=m\delta^{\dot\alpha}_{\ \dot\beta}\,.
    \end{aligned}
\end{equation}
Equivalently, the four-component spin sums are
\begin{equation}
    u^I(p)\bar u_I(p)=\slashed p+m,\qquad
    v^I(p)\bar v_I(p)=-\slashed p+m\,.
\end{equation}
For an internal fermion in $d$ dimensions, the same spin-sum structure is applied before the loop momentum is split into four- and extra-dimensional parts.
Compared with the four-dimensional numerator, the required additional term is the extra-dimensional Clifford insertion:
\begin{equation}
    \begin{aligned}
    \sum_I u^I(\bar l,m)\bar u_I(\bar l,m)
    &=
    \slashed{\bar l}+m
    =
    \slashed l+\tilde{\slashed{\mu}}+m\,,
    \\
    \sum_I v^I(\bar l,m)\bar v_I(\bar l,m)
    &=
    -\slashed{\bar l}+m
    =
    -\slashed l-\tilde{\slashed{\mu}}+m\,.
    \end{aligned}
\end{equation}
Here $\tilde{\slashed{\mu}}\equiv\gamma_{[-2\epsilon]}\cdot\vec{\tilde\mu}$ denotes the extra-dimensional Clifford contraction and is kept inside the closed Dirac trace.
The shifted mass $m_*^2=m^2+\tilde\mu^2$ fixes the four-dimensional spinor kinematics; it does not remove this numerator insertion.
For a massless vector, the cut-state sum runs over the two physical helicities; the mass shift changes the four-dimensional momentum used in the spinor representatives and does not introduce an additional polarization state.
For a shifted massless internal vector with $\bar p^\mu=p^\mu+\tilde p^\mu$, $\bar p^2=0$, and $\tilde\mu^2=-\tilde p^2$, the spinor momentum satisfies $p^2=\tilde\mu^2$.
With reference momentum $q$, one may define a null projection
\begin{equation}
    p^{\flat\mu}
    =
    p^\mu-\frac{\tilde\mu^2}{2p\cdot q}\,q^\mu\, ,
\end{equation}
and build the two transverse polarization spinors from $p^\flat$.
Their paired helicity projector is
\begin{equation}
    \begin{aligned}
    \Pi^{\mu\nu}(p,q)
    &\equiv
    \epsilon_{p+}^{\mu}(q)\epsilon_{p-}^{\nu}(q)
    +
    \epsilon_{p-}^{\mu}(q)\epsilon_{p+}^{\nu}(q)
    \\&=
    -\eta^{\mu\nu}
    +
    \frac{p^\mu q^\nu+q^\mu p^\nu}{p\cdot q}
    -
    \frac{\tilde\mu^2 q^\mu q^\nu}{(p\cdot q)^2}\,.
    \end{aligned}
\end{equation}
The last term is a reference-vector term and drops out of gauge-invariant sewings by Ward identities.
In internal vector contractions, the spinor little-group sums use the antisymmetric $SU(2)$ tensors in Eq.~(\ref{eq:completeness}), while closed Lorentz traces are evaluated in the same $d$-dimensional loop algebra as the sewn integrand.
For a physical massive vector line the same separation is simpler: the mass-shifted spinors carry $M_*^2=M^2+\tilde\mu^2$, and the closed $SU(2)$-epsilon and metric contraction gives the $d-2$ term displayed explicitly in Sec.~\ref{sec:massive_vector}.

To illustrate this standard sewing procedure, consider the $s$-channel contribution to the four-photon scattering amplitude in our sQED example.
The partial integrand possessing the $s$-channel branch cut is constructed by sewing two tree-level amplitudes.
The required on-shell tree amplitude is:
\begin{equation} \label{eq:ssgg_4d}
    \begin{aligned}
        \mathcal{A}^{(0)}_{+-\phi \phi^\dagger}(p_1,p_2,p_3,p_4)& = -2ie^2\frac{\bigl(\langle 2 | \sigma^\mu | 1 ] p_{3\mu}\bigr)^2}{s_{13}s_{14}} \\&= 2ie^2\frac{\asbk{231}^2}{s_{12}} \left(\frac{1}{s_{13}} +\frac{1}{s_{14}} \right) \,,
    \end{aligned}
\end{equation}
where $s_{ij} = (p_i+p_j)^2$ and $\asbk{231} \equiv \bra{2} \sigma^\mu |1] p_3^\mu$.
By applying Eq.~(\ref{eq:disc}) and the replacement rule Eq.~(\ref{eq:onshell-to-prop}) purely in 4 dimensions, the sewn integrand is:
\begin{equation} \label{eq:sew-scalar_4d}
    \begin{aligned}
        \eval{\mathcal{A}^{(1)}}_{s}^{\rm cuts, 4D}
    &= \int \frac{d^{4} l}{(2 \pi)^{4}} \frac{4 e^4 \qty(\asbk{2 l_s 1}\asbk{4 l 3})^2}{s_{12}^2 D_{l}^0 D_{l_s}^0 }\\&\qty(
             \frac{1}{D_{p_1 + l}^0} +  \frac{1}{D_{p_1 + l_s}^0}) \qty(\frac{1}{D_{p_3 - l}^0} + \frac{1}{D_{p_3 - l_s}^0})\,,
    \end{aligned}
\end{equation}
where $D_l^m \equiv l^2 - m^2$ and $l_s = -l - p_1 - p_2$.
The numerator can be converted into Dirac traces and further reduced into scalar integrals.

Applying this sewing procedure to the relevant kinematic channels generates a set of partial integrands.
Constructing the correct full integrand from these partial integrands requires resolving two distinct sources of redundancy.

First, as discussed above, replacing the on-shell condition with the full Feynman propagator introduces principal value parts.
These principal value components cause the constructed partial integrand to contain branch cuts associated with kinematic invariants other than the target $s_\mathcal{I}$-channel.
To address this, we use a channel-selection prescription denoted by $\mathcal{P}_{s_k}$.
It is not an independent operator acting on an unreduced integrand.
Operationally, the projection is defined by the following four steps: fix a routing convention for each parent topology; tensor-reduce every channel to the same scalar master basis; attach each parent scalar integral and its rational descendant to a cut set before the $\epsilon$ expansion; and then apply either ordered overlap removal or symmetric weighting.
With this convention, one first decomposes the sewn expression into a scalar-integral basis,
\begin{equation}
    \eval{\mathcal{A}^{(1)}}^{\rm cuts}_{s_k}
    =\sum_\alpha C_\alpha I_\alpha + R\,,
\end{equation}
and then retains only those scalar integrals $I_\alpha$ whose discontinuity contains the $s_k$ channel, together with the associated rational terms generated by the same $d$-dimensional reduction.
Terms with no $s_k$ discontinuity are discarded in that channel.

For reproducibility, we assign to each scalar integral a channel-support set
\begin{equation}
    \mathcal{N}(I_\alpha)=
    \left\{s_k\,\middle|\,\operatorname{Disc}_{s_k} I_\alpha \neq 0 \right\}.
\end{equation}
Here $C_\alpha$ denotes the coefficient multiplying the scalar integral, whereas $\mathcal{N}(I_\alpha)$ denotes the set of cut channels carried by that parent integral.
The factor $|\mathcal{N}(I_\alpha)|$ is therefore only the number of channels over which the same parent scalar integral is shared.
After tensor reduction, the weighted channel prescription acts term by term as
\begin{equation}\label{eq:weighted-projection}
    \mathcal{P}_{s_k}^{\rm weighted}\!\left[C_\alpha I_\alpha\right]=
    \begin{cases}
    \dfrac{1}{|\mathcal{N}(I_\alpha)|}\,C_\alpha I_\alpha, & s_k\in \mathcal{N}(I_\alpha),\\[2mm]
    0, & s_k\notin \mathcal{N}(I_\alpha).
    \end{cases}
\end{equation}
For a parent scalar integral appearing in several cut channels, the symmetric prescription gives total weight
\begin{equation}
    \sum_{s_k\in\mathcal{N}(I_\alpha)}
    \frac{1}{|\mathcal{N}(I_\alpha)|}=1\,,
\end{equation}
so each identified parent scalar integral is counted once after all channels are summed.
All projection operations in this paper are applied only after tensor reduction to a common scalar-integral basis.
The symbol $\mathcal{P}_{s_k}$ is therefore a convention for selecting reduced scalar-basis terms, not an operator on an unreduced loop integrand.
When the same topology appears in different channels, identical parent scalar integrals are identified by their ordered denominator mass pattern together with the external channel labels that define the momentum routing.
If a different routing convention is chosen, it has to be used consistently before assigning the weights.
Rational pieces inherit the channel assignment of their parent $d$-dimensional scalar coefficients before taking the $\epsilon\to0$ limit; in practice, a rational term generated by $C_\alpha^{(d)}I_\alpha^{(d)}$ is assigned the same channel-support set $\mathcal{N}(I_\alpha)$.
If a collected rational remainder receives contributions from several masters, it must be kept as the sum of the parent-attached pieces until after the cut sets have been assigned; assigning an unstructured final remainder to channels is not part of the prescription.
For example, if a reduced parent box $C_4^{(st)}I_4(s,t)$ has both $s$- and $t$-channel discontinuities, then $\mathcal{N}(I_4(s,t))=\{s,t\}$ and the symmetric prescription assigns $1/2$ of this parent term to each channel.
Any integral or rational term generated by the same $d$-dimensional parent coefficient carries the same support set.

Second, higher-point topologies, such as box integrals, intrinsically possess multiple physical branch cuts (e.g., simultaneously in the $s$- and $t$-channels).
Consequently, even after the projection $\mathcal{P}_{s_k}$, these multi-scale integrals will be repeatedly generated when evaluating cuts in different channels, leading to double-counting.
To eliminate this overlap across different channels, we subtract the components that have already been accounted for in previously merged channels or, equivalently, assign a fixed weight to scalar integrals appearing in several channels.

Combining these two resolution steps, the abstract merging process to obtain the full integrand is given by~\cite{bernTwoLoopGgSplitting2004}:
\begin{equation} \label{eq:merging_abstract}
   \mathcal{A}^{(1)}_{\text{full}} = \sum_{k=1}^{n_{\rm ch}} \prod_{j=1}^{k-1} \left(1 - \mathcal{P}_{s_j} \right) \mathcal{P}_{s_k} \qty[ \eval{\mathcal{A}^{(1)}}^{\rm cuts}_{s_k}],
\end{equation}
where $n_{\rm ch}$ is the total number of independent channels.
The inner prescription $\mathcal{P}_{s_k}$ removes scalar-integral components without the target discontinuity, while the outer product $\prod_{j=1}^{k-1} (1 - \mathcal{P}_{s_j})$ implements overlap removal for higher-point topologies.
The ordered expression and the symmetric weighting give the same full answer when all channels have been reduced to the same basis and the identical parent scalar integrals have been identified by the routing convention above.
The explicit examples below use the symmetric weighting of Eq.~(\ref{eq:weighted-projection}); for example, a box integral appearing in two channels is assigned weight $1/2$ in each channel.

Executing these projections requires identifying the singularity structure after tensor reduction.
The integrands are decomposed into standard scalar integral bases using tensor reduction algorithms, for instance the routines implemented in FeynCalc~\cite{shtabovenkoFeynHelpersConnectingFeynCalc2017,shtabovenkoNewDevelopmentsFeynCalc2016}.
Returning to our sQED example, one then evaluates the $t$-channel and $u$-channel cuts.
In pure 4-dimensional sewing, however, the tree-level amplitudes entering the $t$-channel cut, $\mathcal{A}^{(0)}(\gamma^+, \gamma^+, \phi, \phi^*)$ and $\mathcal{A}^{(0)}(\gamma^-, \gamma^-, \phi, \phi^*)$, vanish identically for massless scalars by helicity selection rules.
Consequently, the pure 4-dimensional cut yields zero for these channels.
This exposes a limitation of strict four-dimensional sewing: while it captures the logarithmic and polylogarithmic branch-cut structures, it misses the rational terms $R$ generated by the extra-dimensional loop momentum.
In the next subsection, we include these terms by promoting the internal momenta to $d$ dimensions.

\subsection{Recovering rational terms via mass shift}
\label{sec:rational-recovering}

For one-loop EFT matching, rational terms $R$ can contribute to local Wilson coefficients, so a general on-shell reconstruction should keep them.
This does not mean that every four-dimensional matching calculation misses a contribution: in many cases the relevant rational pieces are absent by symmetry or pole structure, lie outside the EFT basis or order being extracted, or are absorbed into the chosen local subtraction convention.
Appendix~\ref{App:rational} explains this point for the examples considered here, including why Ref.~\cite{delleroseWilsonCoefficientsNatural2022} does not require an explicit rational-term contribution to the displayed coefficient.
The mass-shift prescription below is therefore a conservative reconstruction device rather than a claim that rational terms always modify the matched Wilson coefficient.
The massive-vector example in Sec.~\ref{sec:massive_vector} then gives an explicit case in which the retained rational term does enter the local Wilson coefficient.
Pure four-dimensional on-shell sewing omits the pieces generated by the $(-2\epsilon)$-dimensional loop momentum.
In this section, we explain how promoting the sewing procedure to $d=4-2\epsilon$ dimensions recovers these contributions in the 't Hooft--Veltman convention specified in Sec.~\ref{sec:algorithm-assumptions}.

\subsubsection{The origin of rational terms and the mass-shift prescription}

Rational terms originate from the interplay between ultraviolet $(1/\epsilon)$ divergences and $\mathcal{O}(\epsilon)$ terms generated by $d$-dimensional tensor contractions in dimensional regularization~\cite{bernComputationLoopAmplitudes1992}.
Consider a generic $d$-dimensional one-loop amplitude parameterized as $\mathcal{A}^{(1,d)} = \sum_n C_n^{(d)} I_n^{(d)}$.
The scalar integrals $I_n^{(d)}$ can contain $1/\epsilon$ poles.
Meanwhile, the kinematic coefficients $C_n^{(d)}$, which arise from contracting the loop momentum against external $4$-dimensional quantities, carry explicit $d$-dependence.
Taylor expanding the coefficient around $\epsilon = 0$ gives $C_n^{(d)} = C_n^{(4)} + \epsilon \mathcal{R}_n + \mathcal{O}(\epsilon^2)$.
When multiplying this by a divergent integral, the $\epsilon$ pole is cancelled, leaving behind a finite rational piece:
\begin{equation}\label{eq:rational-origin}
    \begin{aligned}
        \lim_{\epsilon \to 0} C_n^{(d)} I_n^{(d)} &= \lim_{\epsilon \to 0} \qty[C_n^{(4)} + \epsilon \mathcal{R}_n + \dots] \qty[\frac{1}{\epsilon} + \text{finite}]
     \\&= \lim_{\epsilon \to 0} C_n^{(4)} I_n^{(d)} + \mathcal{R}_n \,.
    \end{aligned}
\end{equation}
This illustrates how rational terms arise from the extra-dimensional components of the loop momentum.
If one applies strict $d=4$ unitarity cuts, the loop momentum is restricted to 4 dimensions, the $\epsilon \mathcal{R}_n$ piece in the coefficient is artificially set to zero, and the rational term is lost.

To recover these missing terms within the same on-shell sewing setup, we treat the loop momenta as $d$-dimensional vectors.
We isolate the extra-dimensional component by parameterizing the $d$-dimensional loop momentum $\bar{l}$ as
\begin{equation}\label{eq:l-to-landmu}
\bar{l} = l + \vec{\tilde\mu} \,,
\end{equation}
where $l$ is the strictly four-dimensional component, and $\vec{\tilde\mu}$ is the $(-2\epsilon)$-dimensional component.
By definition, they are orthogonal: $l \cdot \vec{\tilde\mu} = 0$.
Consequently, the $d$-dimensional on-shell condition $\bar{l}^2 = m^2$ translates to
\begin{equation}
    l^2 - \tilde\mu^2 = m^2 \quad \implies \quad l^2 = m^2 + \tilde\mu^2 = m_*^2\,.
\end{equation}
From the four-dimensional viewpoint, promoting the cut loop momentum to $d$ dimensions is implemented by the shifted mass $m_*^2=m^2+\tilde\mu^2$~\cite{bernMassiveLoopAmplitudes1996,anastasiouDdimensionalUnitarityCut2007}.
Therefore, in the present prescription, one constructs the tree-level amplitudes for the cut loop legs using massive spinor variables with this shifted mass.
The corresponding $4$-dimensional spinor variables for the loop legs satisfy the mass-shifted Dirac equations of motion:
\begin{equation}\label{eq:EoM-d-dim-l}
\slashed{l}|l^I] = m_*|l^I \rangle \,, \quad
\slashed{l}|l^I\rangle = m_*|l^I]\,,
\qquad
m_*^2=m^2+\tilde\mu^2\,.
\end{equation}
Here $I$ denotes the massive little-group index~\cite{arkani-hamedScatteringAmplitudesAll2021,ochirov2018massiveqcd}.
The extra-dimensional dependence enters through the shifted mass $m_*$ and through closed loop-state traces evaluated in the BMHV algebra.
If the internal particle is massless ($m=0$), the shifted mass is $m_*^2=\tilde\mu^2$, so the cut leg is represented by massive spinor-helicity variables during the sewing process.
A practical implementation of this $\tilde{\mu}$-prescription for tree inputs is given in Appendix~\ref{sec:appendix-recover-mu2-guide}.

The mass shift $m^2 \to m^2+\tilde\mu^2$ is applied only to loop-derived quantities that arise from the $d$-dimensional on-shell conditions, spin sums, polarization sums, gamma traces, or loop-momentum contractions of cut internal states.
For propagator denominators this follows directly from $1/(\bar l^2-m^2)$.
For numerator factors it is the same statement applied to the $d$-dimensional internal algebra before the sewn object is reduced to scalar integrals.
It is not an independent shift of external masses, Lagrangian parameters, EFT Wilson coefficients, or couplings.
The same rule fixes the $\tilde\mu^2$ terms that accompany loop-derived numerator factors in the examples considered here.
\subsubsection{sQED rational term}
\label{sec:sqed}

We return to the four-photon scattering example in massless sQED.
The relevant Lagrangian is $\mathcal{L}_{\text{sQED}} = - \frac{1}{4} F^{\mu\nu}F_{\mu\nu} + (D_\mu \phi)^\dagger D^\mu \phi$.
In the previous subsection, we found that purely 4-dimensional sewing failed to capture the $t$- and $u$-channel contributions because the corresponding tree-level amplitudes vanished.
We now apply our $d$-dimensional mass-shift prescription.

Because the internal scalar $\phi$ is massless, the mass-shifted sewing uses $m_*^2=\tilde{\mu}^2$ for the cut scalar legs.
Consequently, the $t$-channel cut requires sewing two tree-level amplitudes: $\mathcal{A}^{(0)}(\gamma^+, \gamma^+, \phi, \phi^*)$ and its conjugate.
Unlike the strictly massless case, these amplitudes no longer vanish for this shifted-mass kinematics.
Constructing them with the mass-shifted on-shell conditions yields:
\begin{widetext}
\begin{equation}\label{eq:SQED-pp00}
    \begin{aligned}
    \mathcal{A}_{++00^*}^{(\tilde\mu^2)} (p_1,p_3, l, l_t)
    &= - \frac{2[13]^2\tilde\mu^2}{t}
    \qty(\frac{1}{D_{\bar l +p_1}^0 }
    + \frac{1}{D_{\bar{l}_t +p_1} ^0}) \,, \\
    \mathcal{A}^{(\tilde\mu^2)}_{--00^*} (p_2,p_4,- l_t,- l)
    &= \frac{2\langle 24 \rangle^2\tilde\mu^2}{t}
    \qty(\frac{1}{D_{-\bar l +p_2}^0 }
    + \frac{1}{D_{-\bar{l}_t +p_2} ^0}) \,,
    \end{aligned}
\end{equation}
\end{widetext}
where $l_t = -l - p_1 - p_3$.
Notice that these tree amplitudes are explicitly proportional to $\tilde{\mu}^2$.

We proceed to sew these amplitudes following Eq.~(\ref{eq:disc}).
After evaluating the internal contractions and reducing the result to scalar bubbles, the $t$-channel contribution is
\begin{equation}\label{eq:sqed-t-reduced-bubble}
  \begin{aligned}
  \mathcal{P}_t
  \!\left[
    \left.\mathcal{A}^{(1)}_{t}\right|_{\rm cuts}
  \right]
  &=
  \frac{4[13]^2\langle24\rangle^2}{(2\pi)^4t^2}\,
  \epsilon
  \\
  &\times
  \left[
    \frac{s}{6u}I_2(s)
    -\frac{t^2}{6su}I_2(t)
    +\frac{u}{6s}I_2(u)
  \right] .
  \end{aligned}
\end{equation}
The middle term combines the two parent contributions proportional to $I_2(t)$, using $s+t+u=0$.
With the scalar-integral normalization used in Appendix~\ref{sec:appendix-scalar-integral-convention}, $\epsilon I_2(s_i)\to1$ as $\epsilon\to0$, so the bracket in Eq.~(\ref{eq:sqed-t-reduced-bubble}) gives
\begin{equation}
    \frac{s}{6u}-\frac{t^2}{6su}+\frac{u}{6s}=-\frac{1}{3}.
\end{equation}
After the channel weights are assigned in the merging prescription, the rational contribution can be written compactly as
\begin{equation}\label{eq:sqed-t-rational-final}
  \mathcal{P}^{\rm weighted}_t
  \!\left[
    \left.\mathcal{A}^{(1)}_{t}\right|_{\rm cuts}
  \right]
  =
  -\frac{4i}{3(4\pi)^2}
  \frac{[13]^2}{[24]^2}.
\end{equation}
For the channel sum, the rational terms are kept attached to the scalar-basis terms from which they originate, as in the weighted prescription of Eq.~(\ref{eq:weighted-projection}).
We define $\mathcal{R}^x$ by separating the weighted $x$-channel contribution into the ordinary scalar integrals carrying the $x$-channel cut and a finite rational remainder.
The cut-constructible part of the $s$-channel has already been fixed by the four-dimensional sewing calculation; the remaining contribution needed for the rational channel sum is
\begin{equation}
    \mathcal{R}^{s}
    =
    -\frac{i[13]^2}{(4\pi)^2t^2[24]^2}
    \left(24tu+\frac{28}{3}t^2\right).
\end{equation}
This is obtained from the finite limits of the parent-attached dimension-shifted triangle and box contributions, together with $\lim_{\epsilon\to0}\epsilon I_2(s)=1$ and the channel weights assigned before taking the rational limit.
Equation~(\ref{eq:sqed-t-rational-final}) gives
\begin{equation}
    \mathcal{R}^{t}
    =
    -\frac{4i}{3(4\pi)^2}
    \frac{[13]^2}{[24]^2}.
\end{equation}
With the external helicity ordering fixed, the identical-particle crossing that exchanges the two same-helicity photon pairs sends $s\leftrightarrow u$ while keeping the spinor prefactor in the same form.
Therefore
\begin{equation}
    \mathcal{R}^{u}
    =
    \left.\mathcal{R}^{s}\right|_{s\leftrightarrow u}
    =
    -\frac{i[13]^2}{(4\pi)^2t^2[24]^2}
    \left(24ts+\frac{28}{3}t^2\right).
\end{equation}
Using $s+t+u=0$,
\begin{equation}
    \mathcal{R}^{s}+\mathcal{R}^{u}
    =
    \frac{16i}{3(4\pi)^2}
    \frac{[13]^2}{[24]^2}.
\end{equation}
Adding the $t$-channel contribution gives the total rational term
\begin{equation}
    \mathcal{R}_{\rm sQED}
    =
    \mathcal{R}^{s}+\mathcal{R}^{t}+\mathcal{R}^{u}
    =
    \frac{i}{4\pi^2}\frac{[13]^2}{[24]^2}.
\end{equation}
The total rational term therefore comes from the weighted channel sum, not from multiplying the displayed $t$-channel contribution by the number of channels.
Together with the cut-constructible terms, this reproduces the known one-loop amplitude~\cite{anastasiouDdimensionalUnitarityCut2007}.
This example shows that the $d$-dimensional mass-shift prescription reproduces the rational terms that are absent in strict four-dimensional cuts within the same sewing calculation.
\subsection{On-shell-like KIB and tadpole subtraction}
\label{sec:decoupling}

In the preceding subsections, we described a $d$-dimensional sewing procedure that reconstructs the branch-cut dependent scalar integrals ($C_n I_n$ for $n \ge 2$) and the associated rational terms $R$ in the scheme used here.
Evaluating the general one-loop amplitude decomposition in Eq.~(\ref{eq:scalar-integrals-basis-d4}) also reveals two classes of scalar integrals that lack kinematic branch cuts: tadpoles ($I_1$) and kinematically independent bubbles (KIBs, denoted as $I_2(p_i^2 = m_i^2)$).
Because these integrals are independent of the Mandelstam variables governing the scattering process, they possess no discontinuity across any physical channel and are not determined by ordinary unitarity cuts.
Their treatment is therefore part of the renormalization prescription rather than a consequence of the cut construction alone.

For this sector we use an on-shell-like KIB subtraction scheme.
For massless external legs this removes tadpoles $I_1(M^2)$ and replaces the heavy bubble by the subtracted combination
\begin{equation}
    I_2(p^2;M^2,M^2)
    \longrightarrow
    I_2(p^2;M^2,M^2)-I_2(0;M^2,M^2),
\end{equation}
with tadpoles absorbed into local counterterms in the same convention.
In a renormalizable UV theory, tadpoles and KIBs originate as divergent constants within two-, three-, and four-point one-particle-irreducible (1PI) correlation functions.
Because they carry no nontrivial singularity structures, such as logarithms or poles in $s_{ij}$, they shift local parameters of the theory.
To specify how these local constants are removed from the matching calculation, we impose the subtraction conditions below.

To illustrate this mechanism, consider the generic renormalized two-point 1PI function for a light field coupled to a heavy resonance of mass $M$:
\begin{equation}
\Sigma_r(p^2, M^2) = \Sigma(p^2, M^2) + \delta m + p^2\delta Z \,,
\end{equation}
where $\Sigma(p^2, M^2)$ is the bare self-energy, and $\delta m, \delta Z$ are the mass and wavefunction counterterms.
The bare self-energy typically evaluates to a combination of a kinematically dependent bubble and a tadpole:
\begin{equation}
    \Sigma(p^2, M^2) = (a_0 + a_1 p^2) I_2(p^2, M^2) + b_0 I_1(M^2).
\end{equation}
We define the counterterms to absorb the tadpole together with the KIB limit of the bubble, $I_2(0, M^2)$.
Rather than the minimal subtraction ($\overline{\text{MS}}$) scheme, we adopt the following subtraction conditions:
\begin{equation}
    -\delta m = a_0 I_2(0, M^2) + b_0 I_1(M^2)\,,\quad
    -\delta Z = a_1 I_2(0, M^2)\,.
\end{equation}
Under this scheme, the renormalized two-point 1PI function is written in terms of the subtracted bubble combination:
\begin{equation}
    \Sigma_r(p^2, M^2) = (a_0 + a_1 p^2)\bigl[I_2(p^2, M^2) - I_2(0, M^2)\bigr]\,.
\end{equation}
This condition sets the local KIB contribution to the self-energy to zero at the on-shell point, $\Sigma_r(0,M^2)=0$ for a massless external leg.
It is on-shell-like, but it is not identical to the standard on-shell renormalization scheme.
A standard on-shell scheme also fixes the pole residue, whereas the subtraction above gives
\begin{equation}
    \left.
    \frac{d\Sigma_r}{dp^2}
    \right|_{p^2=0}
    =
    a_0 I_2'(0;M^2,M^2),
\end{equation}
for the two-point form used above.
Thus the standard on-shell scheme differs by the finite wavefunction counterterm
\begin{equation}
    \delta Z_{\rm OS}-\delta Z_{\rm KIB}
    =
    -a_0 I_2'(0;M^2,M^2).
\end{equation}
This is a finite field redefinition.
In amplitudes with interactions, the same change induces the corresponding finite redefinition of renormalized couplings and other local parameters.

The same convention is used consistently across higher-point 1PI functions.
For instance, the bare three-point vertex correction $\Gamma(p_1^2, p_2^2, p_3^2; M^2)$ generally decomposes into triangles, bubbles on the external legs, and a tadpole:
\begin{equation}
  \Gamma
  =
  \sum_\alpha b_\alpha I_3^\alpha(s_{ij};\vec m_\alpha^{\,2})
  +
  \sum_i a_i I_2(p_i^2,M^2)
  +
  a_0 I_1(M^2)\,.
\end{equation}
If an external momentum $p_i$ is placed on shell ($p_i^2 \to 0$ for massless external states), the corresponding bubble reduces to a KIB, $I_2(0, M^2)$.
By enforcing an analogous local subtraction condition for the vertex counterterm $\delta g$,
\begin{equation}
  -\delta g
  =
  \sum_i a_i I_2(0,M^2)
  +
  a_0 I_1(M^2)\,,
\end{equation}
the renormalized vertex $\Gamma_r = \Gamma + \delta g$ depends exclusively on cut-constructible triangles and the subtracted bubble combination $\bigl[I_2(p_i^2, M^2) - I_2(0, M^2)\bigr]$.
With the coupling defined by this on-shell vertex condition, this is the higher-point analogue of the subtraction above; the remaining finite ambiguity is the usual choice of renormalized field and coupling definitions.

The prescription used in the examples can be stated as the following rule.
First tensor-reduce the sewn result to the common scalar basis.
For each heavy-containing bubble master in threshold matching, keep the coefficient fixed and replace
\begin{equation}
    \begin{aligned}
    C_2(p^2)\,I_2(p^2;m_1^2,m_2^2)
    &\longrightarrow
    C_2(p^2)\bigl[
    I_2(p^2;m_1^2,m_2^2)
    \\
    &\qquad\qquad
    -I_2(p_0^2;m_1^2,m_2^2)
    \bigr] ,
    \end{aligned}
\end{equation}
where $p_0^2$ is the on-shell value, or zero-invariant threshold value, appropriate to the corresponding external channel.
For the massless external legs used below this is $p_0^2=0$, giving the subtraction $I_2(0;M^2,M^2)$ used in Appendix~\ref{sec:appendix-hard-region}.
Tadpoles and on-shell bubble limits are set to zero by local counterterms in the same convention.
The finite pieces removed by this rule are local scheme choices; changing the convention shifts Wilson coefficients by finite local counterterms or finite field and parameter redefinitions.

In summary, the KIBs and tadpoles absent from ordinary cuts are local terms whose finite parts are scheme dependent.
In the subtraction scheme used in this work, they are removed before matching.
Wilson coefficients quoted in another scheme can differ by finite local counterterms or by a redefinition of the low-energy parameters.
\section{On-shell Matching Procedure}
\label{sec:matching_procedure}

Having fixed the $d$-dimensional sewing and subtraction prescriptions, we now convert the reduced UV amplitude into Wilson coefficients.
The conversion is an amplitude-level matching statement: the reduced UV result is expanded at $p\ll M$, its hard-region local polynomial is expressed in the same on-shell contact-amplitude basis that parameterizes the EFT, and the coefficient of each basis element gives the corresponding Wilson coefficient.
This section makes this map explicit and then applies it in the massive-vector example, where the rational term supplies the leading displayed loop-level contribution to the Wilson coefficient.

\subsection{The matching condition and local amplitude projection}
\label{sec:matching_framework}

At fixed external states, renormalization convention, and infrared prescription, EFT matching equates the low-energy expansion of the renormalized UV amplitude with the EFT amplitude order by order in $1/M$ and in the loop expansion~\cite{skibaTASILecturesEffective2010,henningHowUseStandard2016,henningOneloopMatchingRunning2018,cohenFunctionalPrescriptionEFT2021,chalaEfficientOnshellMatching2024}.
Up to one-loop order $\mathcal{O}(\hbar)$, we write this matching condition schematically as
\begin{equation}\label{eq:master_matching}
    \mathcal{A}_{\text{EFT}}^{(0)} + \mathcal{A}_{\text{EFT}}^{(1)} = \mathcal{A}_{\text{UV}}^{(0)} + \mathcal{A}_{\text{UV}}^{(1)} \,,
\end{equation}
where the superscripts $(0)$ and $(1)$ denote tree-level and one-loop contributions, respectively.
The EFT tree amplitude contains the local contact terms whose Wilson coefficients are being fixed, while the EFT loop amplitude is built from lower-order EFT interactions and reproduces the shared infrared behavior.
At one loop, the EFT side is evaluated with the lower-order coefficients $c_i^{(0)}$ and the counterterms defined in the same infrared and renormalization convention.
With this convention fixed, the short-distance coefficient is read from the hard-region local part of the UV loop amplitude.

Wilson coefficients are determined by the short-distance part of this equation.
In the low-energy limit $s_{ij}\ll M^2$, where $s_{ij}$ are external kinematic invariants, the UV amplitude is asymptotically expanded in powers of $1/M$ using the standard hard region expansion~\cite{smirnovAppliedAsymptoticExpansions2002,benekeAsymptoticExpansionFeynman1998,jantzenFoundationGeneralizationExpansion2011}.
This expansion separates the amplitude into a non-local piece containing infrared structures, such as $1/s_{ij}$ poles or $\log(s_{ij})$ branch cuts generated by light states, and a local piece.
The local piece $\mathcal{A}_{\text{UV}}^{\text{local}}$ consists of polynomials in the external momenta and spinor-helicity variables (or polarization vectors), up to the chosen order in the low-energy expansion.

In the modern on-shell amplitude formalism, local contact interactions generated by integrating out heavy physics can be organized in independent local amplitude bases~\cite{shadmiEffectiveFieldTheory2019,Ma2019gtx,henningConstructingEffectiveField2019,durieuxElectroweakEffectiveField2020,durieuxConstructingMassiveOnshell2020,dongConstructingOnshellOperator2021,dongConstructingGenericEffective2022,Liu:2023jbq}.
We denote these bases as $\mathcal{B}_i(\lambda, \tilde{\lambda})$, which are constructed entirely from the external spinor-helicity variables $\lambda$ and $\tilde{\lambda}$ of the light degrees of freedom.
After the low-energy expansion, the local part of the UV amplitude is parameterized as a linear combination of these independent spinor structures, suppressed by powers of $M$:
\begin{equation}
    \begin{aligned}
        \left.\mathcal{A}_{\text{UV}}\right|_{p\ll M}
        &=
        \mathcal{A}_{\text{UV}}^{\text{non-local}}
        +
        \mathcal{A}_{\text{UV}}^{\text{local}} \,,
        \\
        \mathcal{A}_{\text{UV}}^{\text{local}}
        &=
        \sum_{i} \frac{\tilde{c}_i}{M^{d_i-4}} \mathcal{B}_i(\lambda, \tilde{\lambda}) \,.
    \end{aligned}
\end{equation}
Here, $d_i$ is the mass dimension of the corresponding operator, and the coefficients $\tilde{c}_i$ encode the high-energy UV dynamics.
The symbols $\tilde c_i$ denote coefficients in the UV local expansion; after matching they determine the EFT Wilson coefficients $c_i^{(0)}$ and $c_i^{(1)}$ order by order.

On the EFT side, propagation of light fields gives the same infrared/non-local structures, while higher-dimensional local operators $\mathcal{O}_i$ generate contact amplitudes.
The on-shell matrix element of a local operator $\mathcal{O}_i$ is represented by its corresponding amplitude basis element $\mathcal{B}_i$, so we write
\begin{equation}
    \begin{aligned}
        \mathcal{A}_{\text{EFT}} &= \mathcal{A}_{\text{EFT}}^{\text{non-local}} + \sum_i c_i \langle \mathcal{O}_i \rangle \\&= \mathcal{A}_{\text{EFT}}^{\text{non-local}} + \sum_i c_i \mathcal{B}_i(\lambda, \tilde{\lambda}) \,,
    \end{aligned}
\end{equation}
where $c_i$ are the unknown Wilson coefficients to be determined.

With a common infrared prescription, the EFT reproduces the non-local low-energy behavior of the UV theory order by order~\cite{henningOneloopMatchingRunning2018,cohenFunctionalPrescriptionEFT2021,ellisMixedHeavyLightMatching2016}.
The shared non-local terms then cancel in the amplitude difference.
The local matching conditions used below may be written as
\begin{equation}\label{eq:local_matching}
    \begin{aligned}
    \sum_i c_i^{(0)}\mathcal{B}_i
    &=
    \mathcal{A}_{\text{UV}}^{(0),\,\text{local}},
    \\
    \sum_i c_i^{(1)}\mathcal{B}_i
    &=
    \mathcal{A}_{\text{UV}}^{(1),\,\text{hard}} .
    \end{aligned}
\end{equation}
Non-local infrared terms are reproduced by the EFT loops and cancel in the common infrared prescription.
If loops built from lower-order EFT operators generate additional analytic local terms in a different setup, those terms must be treated in the same renormalization convention before reading off $c_i^{(1)}$.

Thus the coefficients $c_i$ are obtained by computing the local hard-region expansion of the UV amplitude in the stated common scheme and projecting the result onto the chosen independent basis $\mathcal{B}_i$~\cite{smirnovAppliedAsymptoticExpansions2002,benekeAsymptoticExpansionFeynman1998,jantzenFoundationGeneralizationExpansion2011,delleroseWilsonCoefficientsNatural2022,chalaEfficientOnshellMatching2024,gerardthooftScalarOneLoop1979}.
Because the on-shell bases $\mathcal{B}_i$ are constructed after quotienting by redundancies associated with equations of motion and integration by parts~\cite{dongConstructingGenericEffective2022,liCompleteSetDimension82021,Liu:2023jbq}, this projection maps directly onto the corresponding EFT contact amplitudes.

The channel-based sewing prescription supplies the cut-constructible terms and rational descendants before the $\epsilon$ expansion.
The matching step then keeps the local hard-region part and projects it onto the basis above.
The following subsection explains how that hard-region projection is performed on the reduced one-loop result.

\subsection{Hard region expansion of the loop integrand}
\label{sec:hard_expansion}

At tree level, the local part $\mathcal{A}_{\text{UV}}^{(0),\,\text{local}}$ is obtained by Taylor expanding heavy-resonance propagators in the small kinematic invariants $s_{ij} \ll M^2$:
\begin{equation}\label{eq:prop-expansion_tree}
    \frac{1}{s_{ij}-M^2} = -\frac{1}{M^2} - \frac{s_{ij}}{M^4} - \frac{s_{ij}^2}{M^6} + \mathcal{O}\left(\frac{1}{M^8}\right) \,.
\end{equation}

At one loop, the hard-region projection is applied before retaining the Wilson coefficient.
We use the method of integration by regions~\cite{smirnovAppliedAsymptoticExpansions2002}, in which the full loop integral is asymptotically expanded by evaluating the integrand in distinct momentum scaling regions.
For matching onto local EFT operators, the only relevant contribution arises from the \textit{hard region}, where the loop momentum $l$ is of the order of the heavy mass $M$, such that $l \sim M \gg p_i, m_i$ (where $p_i$ and $m_i$ are external momenta and light masses).
The \textit{soft region}, where $l \sim p_i$, reproduces the EFT loop contribution in the common infrared prescription and does not change the hard-region coefficient in Eq.~(\ref{eq:local_matching}).

To obtain $\mathcal{A}_{\text{UV}}^{(1),\,\text{local}}$, we apply the hard-region expansion to the $d$-dimensional integrand constructed via the sewing method (Sec.~\ref{sec:framework}) before performing the loop integration.
The expansion is generated by iterating exact algebraic identities for the propagators.
For heavy and light propagators the identities used for this iteration are
\begin{equation}\label{eq:hard-expansion_heavy}
    \frac{1}{(l+p_i)^2-M^2}
    =
    \frac{1}{l^2-M^2}
    \left[
    1-\frac{p_i^2+2l\cdot p_i}{(l+p_i)^2-M^2}
    \right] .
\end{equation}
\begin{equation}\label{eq:hard-expansion_light}
    \frac{1}{(l+p_i)^2-m_i^2}
    =
    \frac{1}{l^2}
    \left[
    1-\frac{p_i^2+2l\cdot p_i-m_i^2}{(l+p_i)^2-m_i^2}
    \right] .
\end{equation}
Iterating these identities gives the hard-region expansion in external momenta and light masses.
After loop integration, the resulting single-scale vacuum integrals reduce to local terms organized as powers of $1/M$.
At the desired order, iterating Eqs.~(\ref{eq:hard-expansion_heavy}) and~(\ref{eq:hard-expansion_light}) leaves propagator denominators only of the form $l^2$ and $l^2-M^2$.
After the hard-region expansion, the sewn integrand has the form
\begin{equation} \label{eq:expanded_integrand_general}
    \sum_{\alpha,\beta,\{i\}}
        \frac{
            \mathcal{N}_{\alpha,\beta}^{\{i\}}(l,p_{\rm ext},\epsilon_{\rm ext})
        }{
            (l^2)^\alpha (l^2-M^2)^\beta
        } .
\end{equation}
All dependence on external momenta, polarizations, and loop momentum that remains in the numerator is collected in the polynomials $\mathcal{N}_{\alpha,\beta}^{\{i\}}$.
Because only a single mass scale $M$ remains in the denominators, the resulting integrals reduce to standard vacuum bubble formulas.
By dimensional analysis, the result of these integrations is a sum of local terms suppressed by powers of the heavy scale, within the truncation order used in the matching calculation.

Notice that terms in Eq.~(\ref{eq:expanded_integrand_general}) with $\beta = 0$ contain no poles in $M^2$.
These correspond to loop diagrams involving only light particles (e.g., pure standard model loops without heavy resonances).
Such diagrams contribute to the infrared (soft) structure common to the UV and EFT descriptions and do not generate local Wilson coefficients at scale $M$ in the hard-region matching.

In the examples, we first tensor-reduce the full sewn integrand to standard scalar integrals, Eq.~(\ref{eq:scalar-integrals-basis-d4}), and then apply the hard-region expansion to the scalar masters.
For instance, the ordered triangle integral $I_3(p^2;0,M^2,0)$, with the middle propagator carrying mass $M$, natively contains complex polylogarithmic branch cuts but evaluates in the hard region strictly to a local polynomial series:
\begin{equation}
    \begin{aligned}
        &I_3(p^2;0,M^2,0)\Big|_{\text{hard}} = \frac{1}{M^2} + \frac{1}{M^2}\log\left(\frac{\mu^2}{M^2}\right) \\&\quad\qquad- \frac{3p^2}{4M^4} - \frac{p^2}{2M^4}\log\left(\frac{\mu^2}{M^2}\right) + \mathcal{O}\left(\frac{p^4}{M^6}\right) \,.
    \end{aligned}
\end{equation}
The expansion isolates the $\mathcal{O}(1/M^n)$ terms.
When multiplied by the kinematic coefficients $C_n$ (which are polynomials in the external momenta and the spinor basis $\mathcal{B}_i$), this generates the local effective operators at the desired order.
The hard-region expansions for the scalar integrals used in the matching examples are collected in Appendix~\ref{sec:appendix-hard-region}.

\subsection{Massive-vector matching example}
\label{sec:wilson-example}
\label{sec:massive_vector}

To illustrate the use of the on-shell matching prescription, we consider a simple model featuring a heavy vector $A_\mu$, a heavy scalar $\phi$, and massless Dirac fermions $\psi$ and $\chi$.
This deliberately simple example displays the sewing prescription and the matching calculation in a setting where the $\tilde\mu$-dependent rational term can be isolated.
Its purpose is to show explicitly how these numerator terms generate the leading displayed local loop contribution after the hard-region expansion, while the external on-shell basis absorbs the finite field-redefinition effects discussed above.
We use this toy model to explain how the rational reconstruction works and how the rational contribution enters the Wilson coefficient, without dealing with a complete gauge theory.
The relevant interactions are given by:
\begin{equation}
    \begin{aligned}
        \mathcal{L} =\,& -\frac{1}{4}F^{\mu\nu}F_{\mu\nu} + \frac{1}{2} M^2 A^\mu A_\mu + \frac{1}{2}\partial_\mu\phi\,\partial^\mu\phi - \frac{1}{2}M^2\phi^2\\[1mm]
        & + i\,\bar{\psi}\gamma^\mu\partial_\mu\psi + i\,\bar{\chi}\gamma^\mu\partial_\mu\chi + \mathcal{L}_{\rm int}\,,\\[1mm]
        \mathcal{L}_{\rm int} =\,& -g\phi\, A_\mu A^\mu - g_1\phi\,\bar\psi\psi - g_2\phi\,\bar\chi\chi\,.
    \end{aligned}
\end{equation}
In this interaction, $g$ has mass dimension one, while $g_1$ and $g_2$ are dimensionless Yukawa couplings.
Thus the loop term proportional to $g^2 g_1 g_2/M^4$ has the same mass dimension as the displayed dimension-six four-fermion coefficient.
Our objective is to use the heavy scalar exchange to fix the leading local four-fermion basis and to use the one-loop massive-vector reconstruction to isolate the rational contribution to the local loop structure.
Specifically, we consider the four-fermion scattering process $\psi^- \bar{\psi}^- \chi^- \bar{\chi}^-$.
Its independent on-shell EFT amplitude basis at dimension-6 ($\mathcal{O}(1/M^2)$) is the spinor structure $\langle 12 \rangle \langle 34 \rangle$.
We use the tree result to fix the leading basis and the loop reconstruction to display the rational local contribution within the same spinor structure.

At tree level, heavy scalar exchange mediates the four-fermion interaction:
\begin{equation}
\begin{tikzpicture}[baseline=.1ex]
  \begin{feynman}
    \vertex (i1) at (-2, 1) {\(\psi^-_1\)};
    \vertex (i2) at (-2,-1) {\(\bar\psi^-_2\)};
    \vertex (v1) at (-1, 0);
    \vertex (v2) at (1, 0);
    \vertex (o1) at ( 2, 1) {\(\chi^-_3\)};
    \vertex (o2) at ( 2,-1) {\(\bar\chi^-_4\)};

    \diagram* {
      (i1) --[fermion] (v1),
      (v1) --[scalar, dashed] (v2),
      (v2) --[fermion] (o2),
      (v1) --[fermion] (i2),
      (o1) --[fermion] (v2),
    };
  \end{feynman}
\end{tikzpicture}
\end{equation}
The corresponding tree-level UV amplitude is
\begin{equation}
\mathcal{A}_{\text{UV}}^{(0)}
= -\frac{i g_1 g_2 \langle 12 \rangle \langle 34 \rangle}{s - M^2}\,.
\end{equation}
Performing the hard region expansion (Taylor expansion in $s/M^2$) yields the local amplitude series:
\begin{equation}\label{eq:example-matching-UVtree}
    \mathcal{A}_{\text{UV}}^{(0),\,\text{local}} = \frac{i g_1 g_2}{M^2}\langle 12 \rangle \langle 34 \rangle + \frac{i g_1 g_2 s}{M^4}\langle 12 \rangle \langle 34 \rangle + \mathcal{O}\left(\frac{1}{M^6}\right)\,.
\end{equation}
The first term corresponds to the dimension-6 basis $\mathcal{B} = \langle 12 \rangle \langle 34 \rangle$.
Applying the matching condition Eq.~(\ref{eq:local_matching}), the tree-level Wilson coefficient is $c^{(0)} = g_1 g_2 / M^2$.

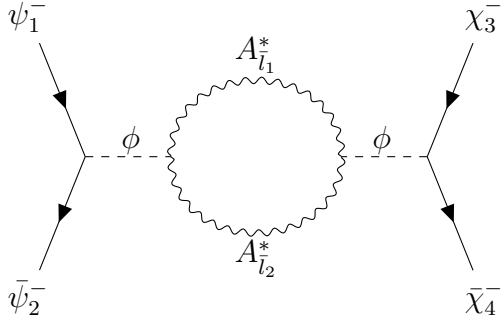
\begin{figure}[tp]
\centering
\begin{tikzpicture}[scale=0.75]
  \begin{feynman}
    % External fermions
    \vertex (i1) at (-4, 2.5) {\(\psi^-_1\)};
    \vertex (i2) at (-4,-2.5) {\(\bar\psi^-_2\)};
    \vertex (o1) at ( 4, 2.5) {\(\chi^-_3\)};
    \vertex (o2) at ( 4,-2.5) {\(\bar\chi^-_4\)};

    % Scalar propagator endpoints
    \vertex (a) at (-3, 0);
    \vertex (b) at ( 3, 0);
    \vertex (c) at (-1.5, 0);
    \vertex (d) at ( 1.5, 0);

    % Loop top and bottom
    \vertex (t) at (0, 1.8) {\(A^*_{\bar{l}_1}\)};
    \vertex (bt) at (0, -1.8) {\(A^*_{\bar{l}_2}\)};

    % Diagram
    \diagram* {
      (i1) -- [fermion] (a) -- [fermion] (i2),
      (o1) -- [fermion] (b) -- [fermion] (o2),
      (a) -- [scalar] (c),
      (b) -- [scalar] (d),
      (c) -- [photon, half left, looseness=1.2, min distance=1.8cm] (d),
      (d) -- [photon, half left, looseness=1.2, min distance=1.8cm] (c),
    };

    \node at (-2.2, 0.3) {\(\phi\)};
    \node at ( 2.2, 0.3) {\(\phi\)};
  \end{feynman}
\end{tikzpicture}
\caption{The one-loop diagram generated by attaching an $A_\mu$ bubble to the scalar propagator $\phi$.}
\label{fig:phi-loop}
\end{figure}

We now compute the one-loop correction to show how the rational term contributes to the Wilson coefficient in the same on-shell basis.
For this helicity configuration, the only non-vanishing unitarity cut occurs in the $s$-channel, where the internal states are two massive vector bosons, $A_\mu$, exchanging a scalar $\phi$ (see Fig.~\ref{fig:phi-loop}).

The relevant tree amplitudes are $\mathcal{A}^{(0)}(\psi^-_1 \bar\psi^-_2 A^*_{\bar l_1} A^*_{\bar l_2})$ and $\mathcal{A}^{(0)}(\chi^-_3 \bar\chi^-_4 A^*_{-\bar{l}_1} A^*_{-\bar l_2})$.
Following the mass-shift prescription detailed in Sec.~\ref{sec:rational-recovering}, the cut vector bosons are represented by four-dimensional massive spinors with shifted mass $M_*^2=M^2+\tilde{\mu}^2$.
The scalar propagator and the final loop denominators still carry the physical mass $M$; the shifted mass appears only in the internal spinor representatives used to reconstruct the numerator.
These four-point tree amplitudes are constructed by sewing the fundamental three-point vertices:
\begin{align}\label{eq:lg-l1l2}
    \mathcal{A}^{(0)}(\psi^-_1 \bar\psi^-_2\phi) &= ig_1\langle 12 \rangle\,, \nonumber \\
    \mathcal{A}^{(0)}(\chi^-_3 \bar\chi^-_4\phi) &= ig_2\langle 34 \rangle\,, \nonumber \\
    \mathcal{A}^{(0)}(\phi A^*_{\bar l_1} A^*_{\bar l_2}) &= -ig\frac{[\mathbf{l}_1\mathbf{l}_2]\langle\mathbf{l}_2 \mathbf{l}_1\rangle}{2(M^2+\tilde{\mu}^2)}\,,
\end{align}
where the bold spinors $\mathbf{l}$ represent massive spinors with symmetrized vector little-group indices and mass $M_*$.
Sewing the internal $\phi$ propagators yields the four-point tree amplitudes:
\begin{align}
   \mathcal{A}^{(0)}(\psi^-_1 \bar\psi^-_2 A^*_{\bar l_1} A^*_{\bar l_2})
   &= \frac{-i g\,g_1\,\langle12\rangle\,[\mathbf{l}_1\mathbf{l}_2]\,\langle \mathbf{l}_2 \mathbf{l}_1\rangle}
             {2\,(M^2+\tilde\mu^2)\,(s - M^2)},\\[1mm]
   \mathcal{A}^{(0)}(\chi^-_3 \bar\chi^-_4 A^*_{-\bar l_1} A^*_{-\bar l_2})
   &= \frac{-i g\,g_2\,\langle34\rangle\,[\mathbf{l}_1\mathbf{l}_2]\,\langle \mathbf{l}_2 \mathbf{l}_1\rangle}
              {2\,(M^2+\tilde\mu^2)\,(s - M^2)}\,.
\end{align}

The one-loop UV integrand is obtained by sewing these two 4-point amplitudes along the massive vector lines.
This requires summing over the internal physical states using the shifted spinor completeness relations stated in Eq.~(\ref{eq:completeness}) and then evaluating the closed vector trace in the BMHV loop algebra.
The little group indices of the cut loop momenta $\bar{l}_1$ and $\bar{l}_2$ in the left-hand tree amplitude contract with those in the right-hand amplitude.
Specifically, the numerator of the sewn integrand contains the squared spinor product:
\begin{equation}
    ([\mathbf{l}_1\mathbf{l}_2]\langle\mathbf{l}_2 \mathbf{l}_1\rangle)^2 \equiv [l_{1\,\{I_1}l_{2\,\{J_1}]\langle l_{2\,J_2\}}l_{1\,I_2\}}\rangle [l^{\{I_1}_1 l^{\{J_1}_2]\langle l^{J_2\}}_2 l^{I_2\}}_1\rangle \,.
\end{equation}
To evaluate this, we apply the mass-shifted completeness relations for the massive spinors.
For a single loop momentum, say $l_1$, the symmetrized vector little-group contraction gives
\begin{equation}
  \begin{aligned}
    &|l_1^{\{I_1}\rangle_\alpha [l_1^{I_2\}}|_{\dot\alpha}
      |l_{1\{I_1}\rangle_\beta [l_{1I_2\}}|_{\dot\beta}
    \\
    &=
    -2 (M^2+\tilde{\mu}^2)\epsilon_{\alpha\beta}\epsilon_{\dot{\alpha}\dot{\beta}}
    +(l_1)_{\beta\dot{\alpha}}(l_1)_{\alpha\dot{\beta}}\,.
  \end{aligned}
\end{equation}
Here $l_1^2=M_*^2=M^2+\tilde\mu^2$, and an analogous identity holds for $l_2$.
The braces denote the same symmetrized vector little-group contraction as in the bold-spinor product above.

Substituting these identities back into the squared spinor product converts the sewn little-group contraction into a closed vector-state contraction.
The open spinor representatives remain four-dimensional, while the metric contraction in the closed internal-state trace is evaluated with the $d$-dimensional loop algebra.
This is why the trace over transverse vector states gives $g_{[d]\,\mu}^{\ \ \mu}-2=d-2$ in the formula below.
Using $l_2=-l_1-(p_1+p_2)$ and $l_1^2=l_2^2=M_*^2$, the spinor contractions reduce to
\begin{equation}
  \begin{aligned}
    &\frac{([\mathbf{l}_1\mathbf{l}_2]\langle\mathbf{l}_2 \mathbf{l}_1\rangle)^2}{4(M^2+\tilde\mu^2)^2} \\
    &= d - 2
    +\left(-1-\frac{l_1\cdot (p_1+p_2)}{M^2+\tilde\mu^2}\right)^2 .
  \end{aligned}
\end{equation}

Inserting this evaluated numerator back into the sewing formula, the $s$-channel partial integrand takes the form:
\begin{equation}
    \begin{aligned}
        \eval{\mathcal{A}^{(1)}}_{s}^{\rm cuts}
        &= g^2 g_1 g_2\,\langle 12 \rangle\langle 34 \rangle
        \int \frac{\dd^d\bar l}{(2\pi)^d}
        \frac{N_s(\bar l)}{D_s(\bar l)}\,,
    \end{aligned}
\end{equation}
where
\begin{equation}
    \begin{aligned}
    N_s(\bar l)
    &=d-2+\qty(-1+\frac{-\bar l \cdot (p_1+p_2)}{M^2})^2,\\
    D_s(\bar l)
    &=(s-M^2)^2(\bar l^2-M^2)
    \bigl((\bar l+p_1+p_2)^2-M^2\bigr).
    \end{aligned}
\end{equation}
In the last line we have rewritten the numerator in the equivalent direct-vector form with physical propagator mass $M$; since $p_1+p_2$ is four-dimensional, $\bar l\cdot(p_1+p_2)=l\cdot(p_1+p_2)$.
This is the same contraction obtained from the massive-vector numerator $-g_{[d]}^{\mu\nu}+\bar l^\mu\bar l^\nu/M^2$, and it is the form used in the scalar-integral reduction below.

At this stage, we employ standard tensor reduction to decompose the integrand into scalar integrals and retain only the components that possess genuine $s$-channel branch cuts using the channel-selection prescription $\mathcal{P}_s$.
The $\epsilon$-dependence in $d=4-2\epsilon$ generates rational terms during this decomposition.
Equivalently, the numerator branch proportional to the scalar bubble has coefficient
\begin{equation}
    C_2(d)
    =
    d-1-\frac{s}{M^2}+\frac{s^2}{4M^4}
    =
    3-2\epsilon-\frac{s}{M^2}+\frac{s^2}{4M^4}\,,
\end{equation}
where polynomial and tadpole pieces have been removed by the channel projection.
The $-2\epsilon$ part multiplies the UV pole of $I_2(s;M^2,M^2)$ and gives the rational term below.
The channel-selected loop amplitude is
\begin{equation}
\begin{aligned}
    &\mathcal{P}_s\!\left[\eval{\mathcal{A}^{(1)}}_{s}^{\rm cuts}\right] \\
    &\quad=
    \frac{ig^2 g_1 g_2\langle 12 \rangle\langle 34 \rangle}{64\pi^2 M^4(s-M^2)^2}
    \bigl(-12 M^4 + 4 M^2 s - s^2\bigr) \\
    &\qquad\times I_2(s;M^2,M^2) + \mathcal{R}\,.
\end{aligned}
\end{equation}
where the rational term $\mathcal{R}$ is explicitly:
\begin{equation} \label{eq:rational}
    \mathcal{R}
    = \frac{2 i g^2 g_1 g_2}{(4\pi)^2 (s - M^2)^2}
    \langle 12 \rangle\langle 34 \rangle\,.
\end{equation}
This exhibits the rational term that would be missed by a strict four-dimensional cut.

Finally, we perform the hard region expansion to extract the matching contribution.
In the subtraction scheme of Sec.~\ref{sec:decoupling}, the divergent scalar bubble $I_2(s;M^2,M^2)$ is renormalized by subtracting its kinematically independent limit $I_2(0;M^2,M^2)$.
Expanding this renormalized combination in the hard region limit ($s \ll M^2$) yields:
\begin{equation}
    \eval{I_2(s;M^2,M^2)-I_2(0;M^2,M^2)}_\text{hard} =\frac{s}{6 M^2}+\mathcal{O}\qty(\frac{s^2}{M^4})\,.
\end{equation}
When multiplied by the prefactor $\sim M^0/M^4$, this scalar integral contribution begins at $\mathcal{O}(M^{-6})$ and therefore does not contribute to the dimension-6 coefficient at the order considered.

Expanding the rational term $\mathcal{R}$ gives:
\begin{equation}
   \mathcal{R}\Big|_{\text{hard}} = \frac{2i g^2 g_1 g_2}{(4\pi)^2 M^4}\,\langle 12 \rangle\langle 34 \rangle + \mathcal{O}\left(\frac{1}{M^6}\right).
\end{equation}
This term has the local scaling of the same dimension-6 basis and gives the one-loop correction
\begin{equation}
    c^{(1)}
    =
    \frac{2g^2 g_1 g_2}{(4\pi)^2M^4}\,.
\end{equation}
Combining the tree-level coefficient $c^{(0)}=g_1g_2/M^2$ with the one-loop correction, the coefficient through one loop is
\begin{equation}
    c
    =
    c^{(0)}+c^{(1)}
    =
    \frac{g_1g_2}{M^2}
    +
    \frac{2g^2 g_1 g_2}{(4\pi)^2M^4}\,.
\end{equation}
In this example the displayed one-loop contribution to the dimension-6 Wilson coefficient is generated by the rational term.

Appendix~\ref{sec:appendix-matching-4Fermion} contains a separate mixed massive--massless four-fermion calculation as a coefficient-level matching example.
We leave the algebra there because it combines channel projection, KIB subtraction, and hard-region expansion in a nontrivial helicity configuration; in that power-counting setup, the rational remainder is retained by the reconstruction but does not change the displayed Wilson coefficient at the order considered.
\section{Conclusion and Outlook}
\label{sec:conclusion}
In this work, we have resolved the tension between rigorous loop-level extraction and the use of non-redundant amplitude bases by organizing standard $d$-dimensional unitarity sewing, mass-shift internal-state reconstruction, hard-region expansion, and on-shell EFT projection into a comprehensive channel-based on-shell framework for one-loop EFT matching.

A central feature of the construction is that the $d$-dimensional reconstruction is made compatible with channel organization and EFT projection.
External states are kept four-dimensional, while the internal loop-state algebra is reconstructed in $d=4-2\epsilon$ dimensions through the mass-shift representation $m^2\to m^2+\tilde{\mu}^2$.
After the double cuts are sewn, all channel contributions are reduced to a common scalar-integral basis before overlap removal.
The rational terms generated by the extra-dimensional numerator algebra are then assigned together with the parent $d$-dimensional scalar coefficients before the $\epsilon\to0$ limit.
This construction evaluates the cut-constructible parts and their associated rational terms within a single, unified structure, thereby eliminating the need for a separate rational-term reconstruction.

This unified treatment is critical because rational terms can directly contribute to matching coefficients.
The explicit examples illustrate the rational reconstruction and demonstrate its role in loop amplitudes.
In the massive-vector model, the rational term provides the leading loop contribution, which is entirely absent in a strict four-dimensional cut construction.
Furthermore, the mixed massive-massless four-fermion calculation shows how rational remainders are carried through the final projection, even when they cancel in the final Wilson coefficient.

To ensure the physical completeness of the matching prescription, we also specified the subtraction treatment for KIBs and tadpoles.
Because tadpoles and kinematically independent bubbles (KIBs) lack kinematic discontinuities, they cannot be captured by unitarity cuts.
To address this, we apply an on-shell-like subtraction scheme that uses a definite choice of field redefinitions and renormalized parameters to absorb these local constants into counterterms before extracting the Wilson coefficients.
Conversion to other finite conventions follows standard scheme transformations, ensuring the resulting coefficients can be translated to other preferred schemes.

A major practical advantage is that the matching prescription is directly compatible with automated computation tools.
The resulting integrands can be processed using standard tensor reduction algorithms (e.g., FeynCalc's TID routines~\cite{shtabovenkoFeynHelpersConnectingFeynCalc2017,shtabovenkoNewDevelopmentsFeynCalc2016}).
Subsequent application of the hard region expansion isolates the local polynomials, which are then projected onto on-shell EFT amplitude bases.
This final projection preserves the central advantage of the on-shell organization: the result is expressed in a basis where gauge, integration-by-parts, and equations-of-motion redundancies have already been cleanly eliminated.

Typical applications of this prescription include gauge-complete non-Abelian or gravitational examples, translation of the resulting coefficients to other finite conventions, and comparisons of the channel-based organization with known matching zeros~\cite{delleroseWilsonCoefficientsNatural2022}.
\begin{acknowledgments}
We are very grateful to Alex Pomarol, Minyuan Jiang, Gauthier Durieux, Kevin Zhang and Xuxiang Li for valuable discussions.
Z.D. is supported by grant PID2023-146686NB-C31.
T.M. is partly supported by the Yan-Gui Talent Introduction Program (grant No. 118900M128), Chinese Academy of Sciences Pioneer Initiative "Talent Introduction Plan", the Fundamental Research Funds for the Central Universities, and National Natural Science Foundation of China Excellent Young Scientists Fund Program (Overseas).
J.S. is supported by the National Key Research and Development Program of China under Grants No.2020YFC2201501 and No.2021YFC2203004, Peking University under startup Grant No.7101302974, the NSFC under Grants No.12025507, No.12150015, No.12450006, and the Key Research Program of Frontier Science of the Chinese Academy of Sciences (CAS) under Grant No. ZDBSLY-7003.
\end{acknowledgments}

\bibliographystyle{apsrev4-2}
\bibliography{export,extra}

\clearpage
\appendix
\onecolumngrid
\section{The Practical Guide for Recovering D-Dimensional Contribution in Tree-Level Amplitudes}\label{sec:appendix-recover-mu2-guide}\label{app:dimensional-conventions}

In dimensional regularization, a $d$-dimensional loop momentum $\bar l$ is naturally split into its four-dimensional component $l$ and a $(-2\epsilon)$-dimensional part $l_{[-2\epsilon]}\equiv \tilde\mu$.
We adopt the convention $\bar l=l+\tilde\mu$ and the $d$-dimensional on-shell condition $\bar l^{2}= l^2-\tilde\mu^2=m^{2}$, consistent with the main paper.

Although $\tilde\mu$ vanishes in the strict four-dimensional limit ($\epsilon\to0$), its remnants are essential to capture the rational term.
In this section, we provide two complementary constructions to recover this contribution.
One is diagrammatic and one is purely on-shell for recovering the $\tilde\mu$ dependence in tree-level amplitudes.
Both constructions are algorithmic and systematic.

\subsection{\texorpdfstring{Diagrammatic construction}{Diagrammatic construction}}
\label{sec:diagrammatic-mu2}
In this subsection, we outline a systematic diagrammatic procedure that recovers the $\tilde\mu$ contribution of tree-level amplitudes within dimensional regularization:
\begin{enumerate}
    \item \textbf{Draw the relevant Feynman graphs.}
    For the process under consideration, enumerate all tree-level $n$-point diagrams and use Feynman rules to obtain these amplitudes.
    \item \textbf{Promote to $d$ dimensions.}
    Promote the loop momentum to higher dimension according to the replacement $l_{i}\mapsto\bar l_{i}$, leaving all external wave-functions unmodified.
    This yields a partially $d$-dimensional amplitude in which $\tilde\mu^{2}$ enters only through the scalar products of the momentum $\bar{l}_i$.
    \item \textbf{Impose the on-shell conditions.}
    Enforce the on-shell condition $\bar l_{i}^{2}=m_{i}^{2}$ and apply the EoMs to the loop spinors, as is collected in Eq.~\ref{eq:EoM-d-dim-l}.
    After these simplifications, the tree-level amplitude takes the schematic form $A_{n}^{(d)}\bigl(\{\lambda_{i},\tilde\lambda_{i}\};\lambda_{l},\tilde\lambda_{l};\tilde\mu\bigr)$, where $\lambda_{l},\tilde\lambda_{l}$ encode the little-group information of the internal states and the external ones in $\lambda_i,\tilde\lambda_i$.
    For tree-level amplitudes, the dependence on momentum $\bar l$ can be converted to the dependence on $\tilde\mu$, $m^2$ and $l\cdot p_{i}$, since only two types of $\bar{l}$ contraction exist: contractions with external momenta reduce to $l\cdot p_{i}$, while $\bar l^{2}$ is replaced by $m^{2}$.
    Furthermore, the EoMs of $l_i$ spinors can reduce $\slashed{\bar{l}}$ to $\tilde{\slashed{\mu}}$.
\end{enumerate}
The above procedure makes the origin of $\tilde\mu$ manifest.
The $\tilde\mu$ in tree-level amplitudes will contribute to the rational term as discussed in Sec.~\ref{sec:rational-recovering}.

\subsection{\texorpdfstring{Purely on-shell construction}{Purely on-shell construction}}
\label{sec:onshell-mu2}
In this subsection, we outline a purely on-shell construction method to restore the $\tilde\mu^{2}$ dependence.
This on-shell construction is equivalent to the former diagrammatic method.
\begin{enumerate}
    \item \textbf{Construct the relevant three-point amplitudes.}
    If states in the 3-point amplitudes involve $d$-dimensional loop momenta when sewing them to construct the higher-point tree amplitudes, their mass should be shifted via $m^2\mapsto m^2+\tilde\mu^{2}$.
    In particular, massless states involving loop momenta should be considered massive with mass of $\tilde\mu$.
    \item \textbf{Sew 3-point amplitudes to $n$-point amplitudes.}
    After sewing the lower-point amplitudes into higher-point amplitudes, the loop momentum factor $|l_{I}]\langle l^{I}|\;\text{or}\;|l_{I}\rangle[l^{I}|$ should be replaced by higher dimensional loop momentum $P_\pm \slashed{\bar l}$.
    The denominators of the propagators involving the loop momenta should be shifted via $m^2\mapsto m^2+\tilde\mu^{2}$ and become $((l+p)^{2}-m^{2}-\tilde\mu^{2})$.
    Equivalently, when written in terms of the four-dimensional projected momentum $l$, the $d$-dimensional denominator $\bar l^2-M^2$ appears as $l^2-M^2-\tilde\mu^2$.
    This is a representation of the $d$-dimensional cut kinematics instead of a change of the physical Lagrangian mass.
   \item \textbf{Eliminate $\slashed{\bar l}$.}
   Apply the spinor EoM relations in Eq.~\ref{eq:EoM-d-dim-l} and split the higher dimensional momentum $\bar l =l +\vec{\tilde \mu}$ to express the tree-level amplitudes in terms of $l$ and $\tilde{\mu}^2$.
   The resulting amplitudes are denoted as $A^{(d)}_{n,\text{pre}}$.
    \item \textbf{Restore the four-dimensional limit.}
    Spurious components may appear in $A^{(d)}_{n,\text{pre}}$, because the shifted massless internal legs are represented by auxiliary massive spinors.
    These components are removed by imposing the four-dimensional limit below.
    The correct amplitude $A_{n}^{(d)}$ must be equal to 4-dimensional amplitude $A_{n}^{(4)}$ when taking the limit of $\tilde\mu\to0$,
    \begin{equation}
        \lim_{\tilde\mu\to0}A_{n}^{(d)}\bigl(\{\lambda_{i},\tilde\lambda_{i}\};\lambda_{l},\tilde\lambda_{l};\tilde\mu\bigr) = A_{n}^{(4)}\bigl(\{\lambda_{i},\tilde\lambda_{i}\};\lambda_{l},\tilde\lambda_{l}\bigr)\,,
    \end{equation}
    We impose this restriction on $A^{(d)}_{n,\text{pre}}$ and obtain the required $A_{n}^{(d)}$ as
    \begin{equation}
        \label{eq:mu2-subtraction}
        \begin{aligned}
            A_{n}^{(d)}(\dots;\tilde\mu)=&A^{(d)}_{n,\text{pre}}(\dots;\tilde\mu)\\&-\bigl[\lim_{\tilde\mu\to0}A^{(d)}_{n,\text{pre}}(\dots;\tilde\mu)-A^{(4)}_{n}(\dots)\bigr].
        \end{aligned}
    \end{equation}
    The subtraction in Eq.~(\ref{eq:mu2-subtraction}) removes only the part that incorrectly survives in the $\tilde\mu\to0$ limit.
    The $\tilde\mu$-dependent terms that vanish at the level of the tree amplitude but can generate finite rational terms after loop integration are retained.
    
\end{enumerate}

\subsection{Summary}
The diagrammatic and on-shell constructions presented above are equivalent.
The former preserves a tight connection to conventional Feynman rules, whereas the latter integrates naturally with modern generalized-unitarity frameworks.
In practice, the on-shell method tends to yield more compact expressions and is more efficient for UV models with high-spin particles such as graviton and non-abelian gauge bosons.
Both constructions can be used in Sec.~\ref{sec:rational-recovering} when we recover the rational term from sewing the tree-level amplitudes.

The dimensional continuation used in this guide assumes that the four-dimensional external-state algebra can be combined with BMHV internal loop momenta and closed internal-state algebra without additional chiral scheme choices~\cite{Gnendiger:2017pys}.
Genuinely chiral structures involving $\gamma_5$ or Levi-Civita tensors require evanescent operators~\cite{FuentesMartin:2022jrf} and finite scheme choices in dimensional regularization.
This is the standard chiral-theory issue; on-shell formulations can express the associated anomaly constraints through locality and unitarity~\cite{Chen:2014eva}.

\subsection{\texorpdfstring{Numerator mass-shift check}{Numerator mass-shift check}}
\label{sec:appendix-proof-numerator-shift}

Following Sec.~\ref{sec:rational-recovering}, we shift only mass factors generated by the projected loop kinematics, such as $l^2=m^2+\tilde{\mu}^2$, $m_*^2$, or an equivalent cut-state relation.
Lagrangian masses, couplings, counterterms, and Wilson coefficients remain the input parameters of the four-dimensional tree amplitude.

A representative two-point function makes this structure explicit.
For a light field coupled to a heavy field of mass $M$, write the heavy-loop one-particle-irreducible self-energy as
\begin{equation}
    \Sigma(p^2, M^2) = C_2(M^2, p^2) I_2(p^2, M^2) + C_1(M^2) I_1(M^2) \,,
\end{equation}
where $C_n$ collect numerator structures and couplings, with the schematic bubble coefficient $C_2(M^2,p^2)=aM^2+bp^2+\cdots$.

For $p^2\ll M^2$, the bubble expands as
\begin{equation}
    I_2(p^2, M^2) = I_2(0, M^2) + p^2 I_2^\prime(0, M^2) + \mathcal{O}\left(\frac{p^4}{M^4}\right) \,,
\end{equation}
with $I_2^\prime(0,M^2)=c/M^2$ for some constant $c$.
The part of the numerator proportional to $aM^2$ therefore gives
\begin{equation}
    (a M^2) I_2(p^2, M^2) = a M^2 I_2(0, M^2) + a c\, p^2 + \mathcal{O}\left(\frac{p^4}{M^2}\right) \,.
\end{equation}
The first term is kinematically independent and belongs to the mass-renormalization part discussed in Sec.~\ref{sec:decoupling}.
The finite kinematic term is the piece tracked by the $d$-dimensional numerator mass shift.

For a mass factor generated by loop-momentum contraction, the convention used in the sewn tree amplitudes is $aM^2\mapsto a(M^2+\tilde{\mu}^2)$.
This produces the associated rational contribution
\begin{equation}
    C_2(M^2+\tilde{\mu}^2, p^2) I_2
    = \cdots + a\tilde{\mu}^2 I_2(p^2,M^2)
    \equiv \cdots + a I_2[\tilde{\mu}^2] \,.
\end{equation}
With the scalar-integral convention collected in Appendix~\ref{sec:appendix-scalar-integral-convention}, $I_2[\tilde{\mu}^2]$ supplies the corresponding finite contribution at order $\epsilon^0$.
Thus, the numerator mass-shift convention associates each rational contribution with the loop-derived mass structure from which it originates, and the same organizing principle extends to higher-point 1PI functions such as $I_3$ and $I_4$.

\section{Example: Matching of Four-Fermion Operators} \label{sec:appendix-matching-4Fermion}
In Sec.~\ref{sec:framework} we presented an example involving exclusively massless particles, whereas Sec.~\ref{sec:wilson-example} focused on a matching problem with only massive internal states.
To demonstrate how to handle mixed massive--massless cuts and to extract the corresponding Wilson coefficients, we study here a toy model consisting of a massive scalar coupled to massless QED through a Yukawa interaction.
The UV Lagrangian is
\begin{equation}
    \mathcal{L} = -\frac{1}{4}F^{\mu\nu}F_{\mu\nu} + i\bar{\psi}\gamma^\mu (\partial_\mu + i e A_\mu)\psi + \frac{1}{2}\partial^\mu\phi\,\partial_\mu\phi - \frac{1}{2}M^2\phi^2 - g\,\phi\,\bar{\psi}\psi.
\end{equation}
where $\psi$ denotes a massless Dirac spinor.
Our goal is to obtain the Wilson coefficient of the helicity-flipping four-fermion operator generated by integrating out the heavy scalar~$\phi$.
We focus on the four-fermion configuration $\psi^+\,\bar\psi^-\,\psi^+\,\bar\psi^-$ whose dimension-six on-shell contact-amplitude basis is $\sbk{13}\abk{24}$.
Since the Yukawa interaction does not generate this helicity configuration at tree level, the relevant matching starts at one loop.
We therefore compute the one-loop amplitude $A^{(1)}(\psi_1^+,\bar\psi_2^-,\psi_3^+,\bar\psi_4^-)$ relevant for the matching.
The one-loop amplitude is reconstructed by unitarity sewing, so we first list the on-shell tree amplitudes entering the $s$-, $t$-, and $u$-channel cuts, since all three two-particle channels contribute.
In each channel, the internal lines are marked with an asterisk ($*$) to track the $-2\epsilon$ components responsible for rational terms.
The required tree amplitudes are
\begin{align}
    \mathcal{A}^{\text{Tree}}(\psi_1^+, \bar{\psi}_2^-, \psi^{*+}_{3}, \bar{\psi}^{*-}_{4})&=-\frac{2ie^2\sbk{13}\abk{42}}{u-\tilde\mu^2}+\frac{2ie^2\asbk{2\tilde\mu 1}\abk{14}\sbk{13}}{(u-\tilde\mu^2)(t-\tilde\mu^2)} \nonumber 
    -\frac{2ie^2\sbk{13}\abk{42}}{s}+\frac{2ie^2\asbk{2\tilde\mu3}\abk{42}\sbk{21}}{s(t-\tilde\mu^2)},
\end{align}
\begin{align}
    \mathcal{A}^{\text{Tree}}(\psi_1^+,\bar{\psi}_2^-, \psi^{*-}_{3}, \bar{\psi}^{*+}_{4})=-\frac{2ie^2\abk{23}\sbk{41}}{s}-\frac{2ie^2\abk{21\tilde\mu3}\sbk{41}}{s(u-\tilde\mu^2)}+\frac{ig^2\abk{23}\sbk{41}}{u-\tilde\mu^2-M^2},
\end{align}
\begin{align}
    \mathcal{A}^{\text{Tree}}(\psi_1^+\psi_2^+\bar{\psi}_3^{*+}\bar{\psi}_4^{*+})=\frac{ig^2\sbk{13}\sbk{24}}{t-\tilde\mu^2-M^2}-\frac{ig^2\sbk{14}\sbk{23}}{u-\tilde\mu^2-M^2},
\end{align}
\begin{align}
    \mathcal{A}^{\text{Tree}}(\bar{\psi}_1^-\bar{\psi}_2^-\psi_3^{*-}\psi_4^{*-})=\frac{ig^2\abk{13}\abk{24}}{t-\tilde\mu^2-M^2}-\frac{ig^2\abk{14}\abk{23}}{u-\tilde\mu^2-M^2},
\end{align}
\begin{align}
    \mathcal{A}^{\text{Tree}}(\psi_1^+, \bar{\psi}_2^-, \phi^*_{3},\phi^*_{4})=ig^2\asbk{231}\qty(\frac{1}{t-\tilde\mu^2}-\frac{1}{u-\tilde\mu^2}),
\end{align}
\begin{align}
    \mathcal{A}^{\text{Tree}}(\psi_1^+, \bar{\psi}_2^-, \gamma^*_{3,+},\phi^*_{4})=\frac{\sqrt{2}ieg\tilde\mu\sbk{13}^2\abk{12}}{t-\tilde\mu^2}\qty(\frac{1}{t-\tilde\mu^2}-\frac{1}{u-\tilde\mu^2}),
\end{align}
\begin{align}
    \mathcal{A}^{\text{Tree}}(\psi_1^+, \bar{\psi}_2^-, \gamma^*_{3,-},\phi^*_{4})=\frac{\sqrt{2}ieg\tilde\mu\abk{23}^2\sbk{12}}{u-\tilde\mu^2}\qty(\frac{1}{t-\tilde\mu^2}-\frac{1}{u-\tilde\mu^2}).
\end{align}

With these tree amplitudes in hand, we reconstruct the one-loop amplitude by sewing the relevant two-particle cuts.
With the chosen external ordering and helicity assignment, the $u$-channel result follows from the $s$-channel result by $s\leftrightarrow u$.
It is therefore sufficient to evaluate the $s$- and $t$-channel sewings.
Sewing the tree amplitudes across the $s$-channel cut gives the partial integrand carrying the $s$-channel discontinuity,
\begin{equation}
  \begin{aligned}
      \eval{\mathcal{A}}_{s}^{\rm cuts} &= \int \!\frac{\dd^d\bar  l}{(2\pi)^d}
      \sum_{h=\pm}\!
      \frac{i\,\mathcal{A}^{\text{Tree}}(\psi_1^+, \bar{\psi}_2^-, \psi^{*h}_{\bar l_1}, \bar{\psi}^{*-h}_{\bar l_2})
            \mathcal{A}^{\text{Tree}}(\psi_3^+, \bar{\psi}_4^-, \psi^{*h}_{-\bar l_2}, \bar{\psi}^{*-h}_{-\bar l_1})}
           {D_{\bar l_1}\,D_{\bar l_2}} \\[2pt]
      &+\ \frac{i\,\mathcal{A}^{\text{Tree}}(\psi_1^+, \bar{\psi}_2^-, \phi^*_{\bar l_1},\phi^*_{\bar l_2})
                  \mathcal{A}^{\text{Tree}}(\psi_3^+, \bar{\psi}_4^-, \phi^*_{-\bar l_1},\phi^*_{-\bar l_2})}
                 {D_{\bar l_1,M}\,D_{\bar l_2,M}} \\[2pt]
      &+\ \frac{i\,\mathcal{A}^{\text{Tree}}(\psi_1^+, \bar{\psi}_2^-, \gamma^*_{\bar l_1,+},\phi^*_{\bar l_2})
                  \mathcal{A}^{\text{Tree}}(\psi_3^+, \bar{\psi}_4^-, \gamma^*_{-\bar l_1,-},\phi^*_{-\bar l_2})}
                 {D_{\bar l_1}\,D_{\bar l_2,M}} ,
  \end{aligned}
\end{equation} 
 where $\bar l_1= \bar l, \bar l_2=-\bar l+p_1+p_2$.
 Similarly, the $t$-channel cut integrand is
\begin{equation}
  \eval{\mathcal{A}}_{t}^{\rm cuts}
  = \int \!\frac{\dd^d\bar  l}{(2\pi)^d}
    \frac{i\,\mathcal{A}^{\text{Tree}}(\psi_1^+\psi_3^+\bar{\psi}_{\bar l_1}^{*+}\bar{\psi}_{\bar l_2}^{*+})
           \mathcal{A}^{\text{Tree}}(\bar{\psi}_2^-\bar{\psi}_4^-\psi_{-\bar l_1}^{*-}\psi_{-\bar l_2}^{*-})}
         {D_{\bar l_1}\,D_{\bar l_2}},
\end{equation}
where $\bar l_1=\bar l, \bar l_2=-\bar l+p_1+p_3$.
After reducing to the scalar-integral basis, we project onto the desired $s/t$ channel support and remove overlaps by the weighted operation $\mathcal{P}_{s/t}^{\rm weighted}$.
The resulting channel-selected contributions are
\begin{equation}
  \begin{aligned}
    \mathcal{P}_s^{\rm weighted}\qty[\eval{\mathcal{A}}_{s}^{\rm cuts}] &=\frac{i\sbk{13}\abk{24}}{(4\pi)^2t^2}\bigl[ \frac{g^2t(4e^2M^2 t+4\epsilon e^2M^2 t - 2e^2st+2\epsilon e^2 st - g^2s^2)}{s^2}I_2(s)\\& + 2g^4t \left ( I_2(s;M^2,M^2)-I_2(0;M^2,M^2) \right )\nonumber\\&+\frac{2g^4M^2 s^3 +4e^2g^2M^4 t^2 - g^4s^3 u}{s^2}I_3(s;0,M^2,0) \\&+ g^4(2M^2 - s)s I_3(s;M^2,0,M^2) - g^4M^4 t I_4(s,t;M^2,0,M^2,0) \\
    &-g^4\frac{2M^4 t +4M^4 u +4M^2 s u - s u^2}{2}I_4(s,u;0,M^2,0,M^2) \\
    &+g^4\frac{2M^4 t +4M^4 u -4M^2 s u + s^2 u}{2}I_4(s,u;M^2,0,M^2,0)\bigr],
  \end{aligned}
\end{equation}
\begin{equation}
  \begin{aligned}
    \mathcal{P}_t^{\rm weighted}\qty[\eval{\mathcal{A}}_{t}^{\rm cuts}]&=\frac{i\sbk{13}\abk{24}}{(4\pi)^2t^2}\bigl[-2g^4t I_2(t)-4g^4M^2 t I_3(t;0,M^2,0) \\
    &-g^4M^4 t(I_4(s,t;M^2,0,M^2,0)+I_4(t,u;0,M^2,0,M^2))\bigr]\,,
  \end{aligned}
\end{equation}
where the definitions of these scalar integrals are collected in Appendix~\ref{sec:appendix-scalar-integral-convention}, and $I_2(p^2;M^2,M^2)$ is renormalized by subtracting $I_2(0;M^2,M^2)$ following the scheme introduced in Sec.~\ref{sec:decoupling}.
Summing over the weighted channel contributions yields the full one-loop amplitude,
\begin{align}
   &\mathcal{A}^{(1)}=\sum_{i=s,t,u}\mathcal{P}_i^{\text{weighted}}\left[\eval{\mathcal{A}}_{i}^{\rm cuts}\right]\nonumber\\&=\frac{i\sbk{13}\abk{24}}{(4\pi)^2t^2}\bigl[ \frac{g^2t(4e^2M^2 t- 2e^2st- g^2s^2)}{s^2}I_2(s)+\frac{g^2t(4e^2M^2 t- 2e^2tu- g^2u^2)}{u^2}I_2(u)\nonumber\\&-2g^4t I_2(t) + 2g^4t\left ( I_2(s;M^2,M^2)-I_2(0;M^2,M^2)\right ) + 2g^4t \left ( I_2(u;M^2,M^2)-I_2(0;M^2,M^2) \right )\nonumber\\& -4g^4M^2 t I_3(t;0,M^2,0)+\frac{2g^4M^2 s^3 +4e^2g^2M^4 t^2 - g^4s^3 u}{s^2}I_3(s;0,M^2,0)\nonumber\\&+\frac{2g^4M^2 u^3 +4e^2g^2M^4 t^2 - g^4u^3 s}{u^2}I_3(u;0,M^2,0) + g^4(2M^2 - s)sI_3(s;M^2,0,M^2) \nonumber\\&+ g^4(2M^2 - u)uI_3(u;M^2,0,M^2) - g^4M^4 t I_4(s,t;M^2,0,M^2,0)- g^4M^4 t I_4(t,u;0,M^2,0,M^2)\nonumber \\
    &-g^4\frac{2M^4 t +4M^4 u +4M^2 s u - s u^2}{2}I_4(s,u;0,M^2,0,M^2) \nonumber\\
    &+g^4\frac{2M^4 t +4M^4 u -4M^2 s u + s^2 u}{2}I_4(s,u;M^2,0,M^2,0)\bigr]+\mathcal{R},
\end{align}
with the rational piece
\begin{equation}
  \mathcal{R}=\frac{ie^2g^2\sbk{13}\abk{24}}{(4\pi)^2t^2}\left(\frac{2t^2(2M^2+s)}{s^2}+\frac{2t^2(2M^2+u)}{u^2}\right).
\end{equation}
The rational term $\mathcal{R}$ does not contribute to the final dimension-six matching coefficient since it contains no hard scale $M$.

The scalar master integrals entering this amplitude are expanded in local terms in inverse powers of the heavy scale $M$ using the hard-region expansion introduced in Sec.~\ref{sec:hard_expansion}; the required expansions are shown in Appendix~\ref{sec:appendix-hard-region}.
Collecting all contributions up to $\mathcal{O}(M^{-2})$, the Wilson coefficient of the dimension-six EFT basis at loop level is read off from the matching condition
\begin{equation}
  \frac{c}{M^2}\sbk{13}\abk{24}=\frac{4e^2g^2}{9M^2(4\pi)^2}\left(11+6\log\qty(\frac{\mu^2}{M^2})\right)\sbk{13}\abk{24}\,,
\end{equation}
Choosing $\mu=M$ minimizes the logarithm.

\section{\texorpdfstring{Scalar-Integral Conventions}{Scalar-Integral Conventions}}\label{sec:appendix-scalar-integral-convention}
We collect the scalar master integrals used in the four-fermion matching calculation of Appendix~\ref{sec:appendix-matching-4Fermion}.
We adopt the following convention: the mass labels in each scalar integral are ordered in the same sequence as the propagators displayed in the denominator.
This convention distinguishes, for example, $I_3(p^2;0,M^2,0)$ from $I_3(p^2;M^2,0,M^2)$.
\begin{equation*}
    D_q^0\equiv q^2+i\varepsilon,\qquad D_q^M\equiv q^2-M^2+i\varepsilon.
\end{equation*}
\begin{align}
    I_2(p^2)
    &=\dfrac{\mu^{4-d}}{i\pi^{d/2} r_\Gamma}
      \int \frac{\dd^d l}{D_{l+p}^0D_l^0} \\
    I_3(s)
    &=\frac{\mu^{4-d}}{i\pi^{d/2} r_\Gamma}
      \int
      \frac{\dd^d l}{D_l^0D_{l+p_1}^0D_{l+p_1+p_2}^0} \nonumber \\
    I_4(s,u)
    &=\frac{\mu^{4-d}}{i\pi^{d/2} r_\Gamma}
      \int
      \frac{\dd^d l}{D_l^0D_{l+p_1}^0D_{l+p_1+p_2}^0D_{l-p_4}^0} \nonumber \\
    I_4(s,t)
    &=\frac{\mu^{4-d}}{i\pi^{d/2} r_\Gamma}
      \int
      \frac{\dd^d l}{D_l^0D_{l+p_1}^0D_{l+p_1+p_2}^0D_{l-p_3}^0} \\
    I_2(p^2;M^2,M^2)
    &=\dfrac{\mu^{4-d}}{i\pi^{d/2} r_\Gamma}
      \int \frac{\dd^d l}{D_{l+p}^MD_l^M} \\
    I_3(p^2;0,M^2,0)
    &=\dfrac{\mu^{4-d}}{i\pi^{d/2} r_\Gamma}
      \int
      \frac{\dd^d l}{D_l^0D_{l+p_1}^MD_{l+p}^0} \\
    I_3(p^2;M^2,0,M^2)
    &=\dfrac{\mu^{4-d}}{i\pi^{d/2} r_\Gamma}
      \int
      \frac{\dd^d l}{D_l^MD_{l+p_1}^0D_{l+p}^M}
    \end{align}
    \begin{align}
    I_4(s,t;M^2,0,M^2,0)
    &=\dfrac{\mu^{4-d}}{i\pi^{d/2} r_\Gamma}
      \int
      \frac{\dd^d l}{D_l^MD_{l+p_1}^0D_{l+p_1+p_2}^MD_{l-p_3}^0} \\
    I_4(s,u;0,M^2,0,M^2)
    &=\dfrac{\mu^{4-d}}{i\pi^{d/2} r_\Gamma}
      \int
      \frac{\dd^d l}{D_l^0D_{l+p_1}^MD_{l+p_1+p_2}^0D_{l-p_4}^M}\\
    I_4(s,u;M^2,0,M^2,0)
    &=\dfrac{\mu^{4-d}}{i\pi^{d/2} r_\Gamma}
      \int
      \frac{\dd^d l}{D_l^MD_{l+p_1}^0D_{l+p_1+p_2}^MD_{l-p_4}^0}\\
    I_4(t,u;0,M^2,0,M^2)
    &=\dfrac{\mu^{4-d}}{i\pi^{d/2} r_\Gamma}
      \int
      \frac{\dd^d l}{D_l^0D_{l+p_1}^MD_{l+p_1+p_3}^0D_{l-p_4}^M}
\end{align}
where $r_\Gamma$ is defined as follows,
\begin{align}
    r_\Gamma \equiv \frac{\Gamma^2(1 - \epsilon) \Gamma(1 + \epsilon)}{\Gamma(1 - 2\epsilon)},
\end{align} 
and the factor $\mu^{4-d}$ restores the canonical mass dimension of the scalar integrals in $d=4-2\epsilon$ dimensions.

\section{\texorpdfstring{Hard-Region Expansion of the Scalar Masters}{Hard-Region Expansion of the Scalar Masters}}\label{sec:appendix-hard-region}
The following expansions use the ordered mass labels of Appendix~\ref{sec:appendix-scalar-integral-convention}; for boxes, $(s_1,s_2)$ follows the argument order of $I_4(s_1,s_2;\cdots)$.
The fully massless bubble is scaleless in the hard region and vanishes in dimensional regularization.
\begingroup \allowdisplaybreaks
\begin{align}
    I_2(p^2)&=0\\
    I_2(p^2;M^2,M^2)-I_2(0;M^2,M^2)
    &=\frac{p^2}{6 M^2}+\mathcal{O}\qty(\frac{p^4}{M^4}) \\
    I_3(p^2;0,M^2,0)
    &=\frac{1}{M^2}+\frac{1}{M^2}\log\qty(\frac{\mu^2}{M^2})
      -\frac{3p^2}{4M^4}-\frac{p^2}{2M^4}\log\qty(\frac{\mu^2}{M^2})\nonumber \\
    &\quad+\frac{11p^4}{18M^6}+\frac{p^4}{3M^6}\log\qty(\frac{\mu^2}{M^2})
      +\mathcal{O}\qty(\frac{p^6}{M^8}) \\
    I_3(p^2;M^2,0,M^2)
    &=-\frac{1}{M^2}-\frac{p^2}{12M^4}
      -\frac{p^4}{90M^6}+\mathcal{O}\qty(\frac{p^6}{M^8})\\
    I_4(s_1,s_2;M^2,0,M^2,0)
    &=-\frac{2}{M^4}-\frac{1}{M^4}\log\qty(\frac{\mu^2}{M^2})\nonumber \\
    &\quad+\frac{s_1}{6 M^6}+\frac{2 s_2}{M^6}
      +\frac{s_2}{M^6}\log\qty(\frac{\mu^2}{M^2})
      +\mathcal{O}\qty(\frac{1}{M^8}) \\
    I_4(s_1,s_2;0,M^2,0,M^2)
    &=-\frac{2}{M^4}-\frac{1}{M^4}\log\qty(\frac{\mu^2}{M^2})\nonumber \\
    &\quad+\frac{s_2}{6 M^6}+\frac{2 s_1}{M^6}
      +\frac{s_1}{M^6}\log\qty(\frac{\mu^2}{M^2})
      +\mathcal{O}\qty(\frac{1}{M^8}) 
\end{align}
\endgroup

\section{How the Rational Parts Contribute to EFT}\label{App:rational}

In this section, we discuss how the rational parts contribute to the EFT bases.
Even though our sewing method includes these parts, it is very useful to tell which bases cannot be generated from the rational terms.
Our conclusion is that the rational parts do not contribute to EFT if the UV massive fields have a $Z_2$ symmetry.
The reason is that after implementing the Passarino and Veltman decomposition to the loop amplitudes (using the Van Neerven-Vermaseren basis to decompose the loop momentum), the mass $M$ of the internal UV fields only appears in the numerator of the rational terms $\mathcal{R}$, except that there are tree-level propagators of UV fields in the loop amplitude.
This is because if there is a tree-level propagator, one can perform the large mass expansion to get local EFT amplitudes, similar to Eq.~(\ref{eq:rational}).
However, since UV fields have a $Z_2$ symmetry, their pole structures are forbidden.

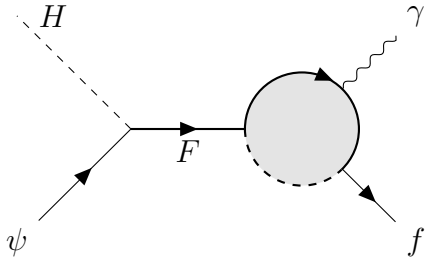
\begin{figure}[ht]
  \centering
  \begin{tikzpicture}[scale=0.75]
    \begin{feynman}
      \vertex (in)    at (-2, -2)    {\(\psi\)};          
      \vertex (v1)    at ( 0,  0);                             
      \vertex (mass)  at (-2,  2);                             
      \vertex (v2)    at ( 3,  0);                  \vertex (v2left)    at ( 2,  0); \vertex (v2rightup)    at ( 3.7071,  0.7071); \vertex (v2rightdown)    at ( 3.7071,  -0.7071);         
      \vertex (out1)  at ( 5,  2) {\(\gamma\)};   
      \vertex (out2)  at ( 5, -2) {\(f\)};   

      \diagram* {
        (in)   -- [fermion] (v1) -- [fermion, thick, edge label'=\(F\)] (v2left),
        (v1)   -- [scalar]  (mass),
        (v2rightup)   -- [photon] (out1),
        (v2rightdown)   -- [fermion] (out2),
      };

      \node[label=right:\(H\)] at (mass) {};

      \begin{scope}
        \def\r{1}
        \fill[gray!30, opacity=0.7] (v2) circle (\r);
        \draw[fermion, thick]
          ($(v2)-(0:\r)$) arc (180:-46:\r);
        \draw[dashed, thick]
          ($(v2)-(0:\r)$) arc (-180:-46:\r);
      \end{scope}
    \end{feynman}
  \end{tikzpicture}
  \caption{3-point loop amplitude induced rational term}
  \label{fig:3pt_rational}
\end{figure}

Next, we will discuss how rational terms contribute to four-point EFT amplitude bases.
As said before if there exist massive poles in the UV amplitudes, rational terms may contribute to EFT bases.
In this case, we can just look at the 3-point UV amplitude that contains the loop, as shown in Fig.~\ref{fig:3pt_rational}.
So if the 3-point loop amplitude does not contain rational terms, there are no rational terms in the full UV amplitude.
For example, the 3-point amplitude for a massless vector ($\gamma$)- a massless fermion ($f$)- a massive fermion ($F$) can never have rational terms, this is because the only rational term can have the form
\begin{equation}
    \mathcal{R} (F f \gamma^+) \sim M \frac{ [1 3] [2 3] }{s_{23}},
\end{equation}
where $M$ is the UV mass scale.

However, this mass scale factor $M$ can never appear in the numerator of the rational term.
This is because if the loop amplitude is from triangle diagrams, the rank of the loop momentum of the integrand should be at most two, and thus the only structure of the numerator for the integrand is
\begin{equation}
    A(F f \gamma^+) \sim \int \frac{\dd^4 l}{(2\pi)^4} \frac{l^\mu l^\nu \epsilon_\mu \bar{\psi} \gamma_\nu \psi }{d_1 d_2 d_3} \,,
\end{equation}
where $d_i$ is the propagator and $\epsilon_\mu$ ($\psi$) is the wave function of external vector (fermion).
Since UV theory is supposed to be renormalizable, the higher-dimensional vector structure in the numerator, such as $ \bar{\psi} \gamma_\nu \gamma_\mu \psi p_3^\mu $, are not allowed.
We find that the structure $\bar{\psi} \gamma_\nu \psi$ in the numerator is forbidden by gauge symmetry, so there are no rational terms in this amplitude.
This is why Ref.~\cite{delleroseWilsonCoefficientsNatural2022} does not consider the rational terms for $g-2$ EFT bases matching.
In Sec.~\ref{sec:wilson-example}, we explicitly show an example in which rational terms indeed contribute to EFT matching.
This is because the loop amplitudes contain the pole structures of the UV massive fields, consistent with the above general discussion.
\end{document}